\documentstyle[aps,psfig,twocolumn]{revtex} 
\begin{document}
\draft
\title{Chaotic enhancement of decay. The effect of classical phase 
space structures}
\author{Yosef Ashkenazy$^1$, Luca Bonci$^1$, Jacob Levitan$^{1,2}$, 
Roberto Roncaglia$^3$}
\address{$^1$ Physics Department, Bar-Ilan University, Ramat-Gan, 
Israel}
\address{$^2$ College of Judea and Samaria, Ariel, Israel} 
\address{$^3$ Piazza S. Salvatore, 1 55100 Lucca, Italy} \date{\today}
\maketitle
\begin{abstract}
We investigate the decay process from a time dependent potential well
in the semiclassical regime.
The classical dynamics is chaotic and the decay rate shows 
an irregular behavior as a function of the system parameters.
By studying the weak-chaos regime we are able to connect the decay
irregularities to the presence of nonlinear resonances in the classical
phase space. A quantitative analytical prediction which
accounts for the numerical results is obtained.
\end{abstract}

\section{introduction}
In the last years, many conjectures has been put forward, and tested 
in various system models, to answer the fundamental question of the 
``quantum chaos problem'': what is the signature of classical chaos 
in the quantum world? Among these, one of the most intriguing is the 
idea according to which the classical chaos can induce large scale 
fluctuations on a genuine quantum phenomenon such as the tunneling 
process. Starting from the seminal paper of Davis and 
Heller~\cite{dh81}, who 
first noted the occurrence of coherent tunneling between regular tori 
separated by a chaotic region, the influence of classical chaos on 
quantum tunneling has been verified in many systems and is now 
accepted in the literature as a fingerprint of classical non 
integrability. It is very simple to describe this effect. Let us 
consider a system which is classically chaotic and invariant under a 
symmetry operation, like for example the space inversion. If the 
classical system supports a regular torus, by symmetry, there might 
be also a second torus which is distinct 
from its symmetric partner, like for instance two symmetric tori 
encircling the bottom of the two wells of a double-well potential. 
Moreover, let us suppose that the two tori are large enough to support 
quantum states. In this condition, the quantum system will show 
coherent tunneling between the states located in the two symmetric 
tori. If now one system parameter is changed (e.g. $\hbar$), contrary 
to the expectations of ordinary semiclassical analysis, the tunneling 
rate shows strong irregularities which can increase or decrease the 
rate by orders of magnitude. This effect does not show up in a system 
whose classical counterpart is integrable.

The tunneling fluctuation is usually interpreted in terms of assisted 
processes, or, using a widespread terminology, as ``Chaos Assisted 
Tunneling'' 
(CAT)~\cite{lb90,hanggi,btu93,tu94,prlnoi,frischat,lgw,prenoi,hanggipreprint}.
An intuitive view of the CAT process is as follows. The presence 
or regular and stochastic motion in the classical phase space 
corresponds, from a quantum point of view, to the possibility of 
having two kind of states: regular ones localized inside the 
symmetric tori and chaotic states, which being extended through the 
chaotic region have a not negligible overlap with regular regions. 
The fluctuations in the tunneling rate are thus explained in terms of 
a three state tunneling process. The quantum particle first tunnels 
from the localized state to an extended chaotic one and then from 
this to the state located in the symmetric torus. 
\begin{figure}[t]
	\centerline{\psfig{figure=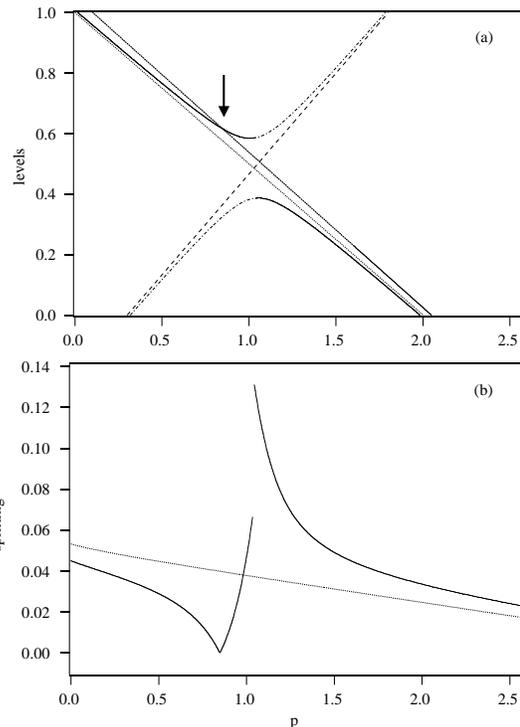,width=3 in}
                   }
	\caption{Sketch of the typical behavior of the energy levels of a 
	classically chaotic quantum system as a parameter $p$ is changed.
        In a) the two solid lines 
	describe a couple of quasi-degenerate levels of different symmetry. 
	The dashed line describes a colliding third level.
	In b) we show the splitting of the two regular levels.
	All the units are arbitrary.}
	\protect\label{fig:split}
\end{figure}

This interpretation is confirmed by the level dynamics of the 
tunneling system. A typical situation is sketched in Fig.~\ref{fig:split} 
where we can observe the change of two quasi-degenerate levels, which 
correspond to the pair of tunneling regular states, as a system 
parameter is varied. In almost the entire parameter range the 
splitting between the two states, and so the direct tunneling 
probability, changes smoothly. However, it may occur that, once the 
parameter is changed, a third level (dashed line) crosses the two 
quasi-degenerate levels. In the non-integrable case states belonging to the same 
symmetry class do not cross each other. Therefore, the appearance of a third 
{\em colliding states} gives rise to an {\em avoided crossing} with 
the state of the doublet which belongs to the same symmetry class. 

The avoided crossing has a twofold consequence on the tunneling process 
under study. First, in the nearby of the crossing we cannot consider 
the tunneling as a process involving only the two quasi-degenerate states.
In this condition the standard two state tunneling becomes a
resonant three state process.
Second, since the colliding third state modifies the energy level of 
only one of the doublet states, the splitting of the two levels 
changes.
Therefore, given the small value of the energy splitting in the semiclassical 
regime, the avoided crossing produces a dramatic modification of 
the levels splitting and consequently of the tunneling rate.
It is important to point out that the rate can increase by 
several orders of magnitude as well as vanish (see the arrow in 
Fig.~\ref{fig:split}) according to the value of the parameters.
Moreover, due to the fact that the energy 
spectrum, in the nonlinear case, does not show any regularity, the 
crossings with a third level do not follow a regular pattern, and the 
overall behavior of the tunneling rate appears to be an irregular 
sequence of peaks~\cite{btu93,tu94,prlnoi,frischat} instead of the 
smooth behavior expected in the regular systems.

These features motivated the widespread idea that classical chaotic 
trajectories can have an active role in the quantum process, helping 
or ``assisting'' the quantum particle to tunnel between the symmetric 
tori. Along this line, path integral techniques have been recently 
used to calculate the contribution to the tunneling stemming from 
complex orbits which connect the symmetric regular tori through the 
classical stochastic layer~\cite{frischat}.

However, a real quantitative theory of the CAT process is still 
lacking. The main reason for this can probably be found in the 
chaotic nature of the third state which prevents any simple 
analytical treatment. Moreover, there are some aspects of the 
phenomenon which do not seem to fit properly with the intuitive 
interpretation given by the CAT picture. For example, the presence of 
strong decreases in the tunneling rate which, together with the 
enhancements, occur as a result of a parameter change, contradicts 
the idea of a tunnel process being ``assisted'' by chaos. Another 
controversial aspect of the effect is the nature of the third state 
which crosses the tunneling doublet. The key point of the CAT picture 
is that the perturbation in the energy splitting is relevant only for 
those crossings involving colliding third states which are located in 
the chaotic region. However, recently, it has been shown that 
similar strong fluctuations can be obtained also in almost integrable 
systems~\cite{prenoi}. In this condition, the third state responsible 
for the fluctuation is by no means chaotic. It is regular and its 
crossing with the tunneling doublet can be directly related to the 
classical phase-space structure and in particular to the destruction 
of the regular tori which are transformed into classical non-linear 
resonances. In particular, a fluctuation (avoided crossing) occurs 
when the energy of the tunneling state corresponds classically to the 
energy of a nonlinear resonance. In other words, in this regime, the 
fluctuations (avoided crossings) are the quantum manifestations of 
the classical ``small denominators'' problem. This discovery also led 
to a first analytical prediction about the positions of the tunneling 
fluctuations which we shall review in 
Section~\ref{sec:semiclassical}. 

In this paper we want to assess whether this picture applies also to 
a different tunneling process, namely the quantum escape of a 
particle which has been initially located inside a potential well. 
From a classical point of view it is clear that the particle can 
overcome the potential barrier of the well only if its energy is 
larger than the barrier height, while in the quantum framework the 
tunneling across the classically forbidden region is always present. 
Clearly this situation is going to be modified when, including the 
ingredient of chaos, we perturb the system by adding a forcing term, 
i.e., a time dependent external force. The perturbation disturbs the 
regular motion of the classical particle and, by increasing its 
energy, makes it possible to the particle to escape over the barrier. 
In the meanwhile, also the quantum process of tunneling changes due 
to the modification of the potential and both the processes 
contribute to the decay~\cite{fendrik}.
Our purpose is to choose a region of the 
system parameters where the classical and the quantum contribution to 
the decay can be separated, in order to study the properties of the 
latter process in connection with the chaotic features of the 
classical phase space. 

By studying a decay process, we shall address a infinite system, 
with a continuous spectrum, and this will prevent us from using 
standard methods, like the diagonalization of the Floquet dynamic 
operator~\cite{floq}, to obtain directly the level splittings responsible
for the tunneling. Anyway we shall be able to analyzes the system by 
resorting to a somewhat simpler method: we shall calculate 
numerically the time evolution of a quantum state initially located 
in the potential well and, by studying the decay of the population in 
the well, we shall be able to obtain the relative strength of the 
tunneling as a function of the system parameters. This will 
allow us to point out the differences between this process of chaos assisted 
decay and the tunneling processes in the presence of chaos.
Finally, we shall be able to show that the picture which singles out 
the classical nonlinear resonances as the main responsible for the 
fluctuations of the tunneling rate, applies also in this 
context, and that the prediction of Ref.~\cite{prenoi}, illustrated in 
Section~\ref{sec:semiclassical}, remains valid. 

\section{The Model: Classical Dynamics}
\label{sec:classical}
\begin{figure}
      \centerline{\psfig{figure=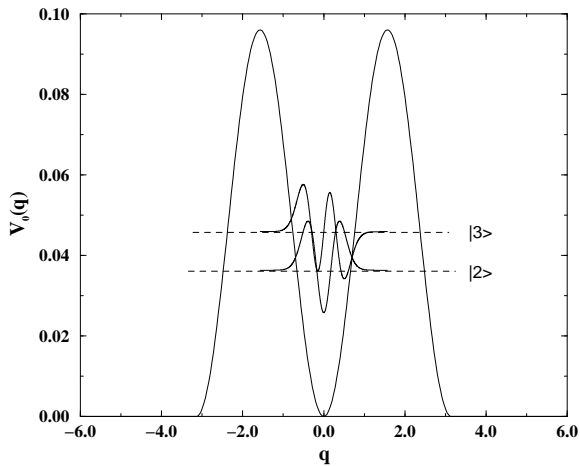,width=3 in}
                 }
      \caption[]{The 
       unperturbed potential of Eq.~\ref{potential}. We also show the wave 
       function of the two states used as initial condition in the numerical 
       calculations.}
      \label{fig:poten}
\end{figure}

In order to analyze the influence of classical chaos on the quantum 
process of escape from a potential well, we introduce a simple 
one-dimensional forced system described by the following Hamiltonian: 
\begin{eqnarray}
\label{hamiltonian}
H &=&\frac{p^{2}}{2}+V(q,t)\\
\label{potential}
V(q,t) &=&V_0 (1-\cos(2q))+ \\
&&\epsilon (1-\cos (2q-\Omega t)) \ \
q=[-\pi,\pi] \\
V(q,t) &=&0 \ \ \ otherwise \ . \nonumber 
\end{eqnarray}
The particle is located initially inside a potential well which has 
the form of a sinusoidal function extended over two periods, as shown 
in Fig.~\ref{fig:poten}, and then it is forced by a time periodic 
perturbation which is considered to be small compared to the static 
potential well, i.e., $\epsilon \ll V_0$. 
We chose a perturbation term which turns system~(\ref{hamiltonian}) 
into a {\em double resonance}-like Hamiltonian.
This choice is dictated by the sake of simplicity. It is indeed clear 
that, as long as we limit our analysis to small perturbations, the 
particular form of the external forcing does not affect the generic 
features of the decay process we want to study. On the other hand, 
the adoption of Hamiltonian~(\ref{hamiltonian}) presents many 
benefits. All the relevant information concerning the dynamic 
properties of our model can be derived from the dynamics of a well 
known system, the {\em double resonance} Hamiltonian which corresponds 
to eq.~(\ref{hamiltonian}) with periodical boundary 
conditions~\cite{r92,ll83}. 

\begin{figure}
      \centerline{\psfig{figure=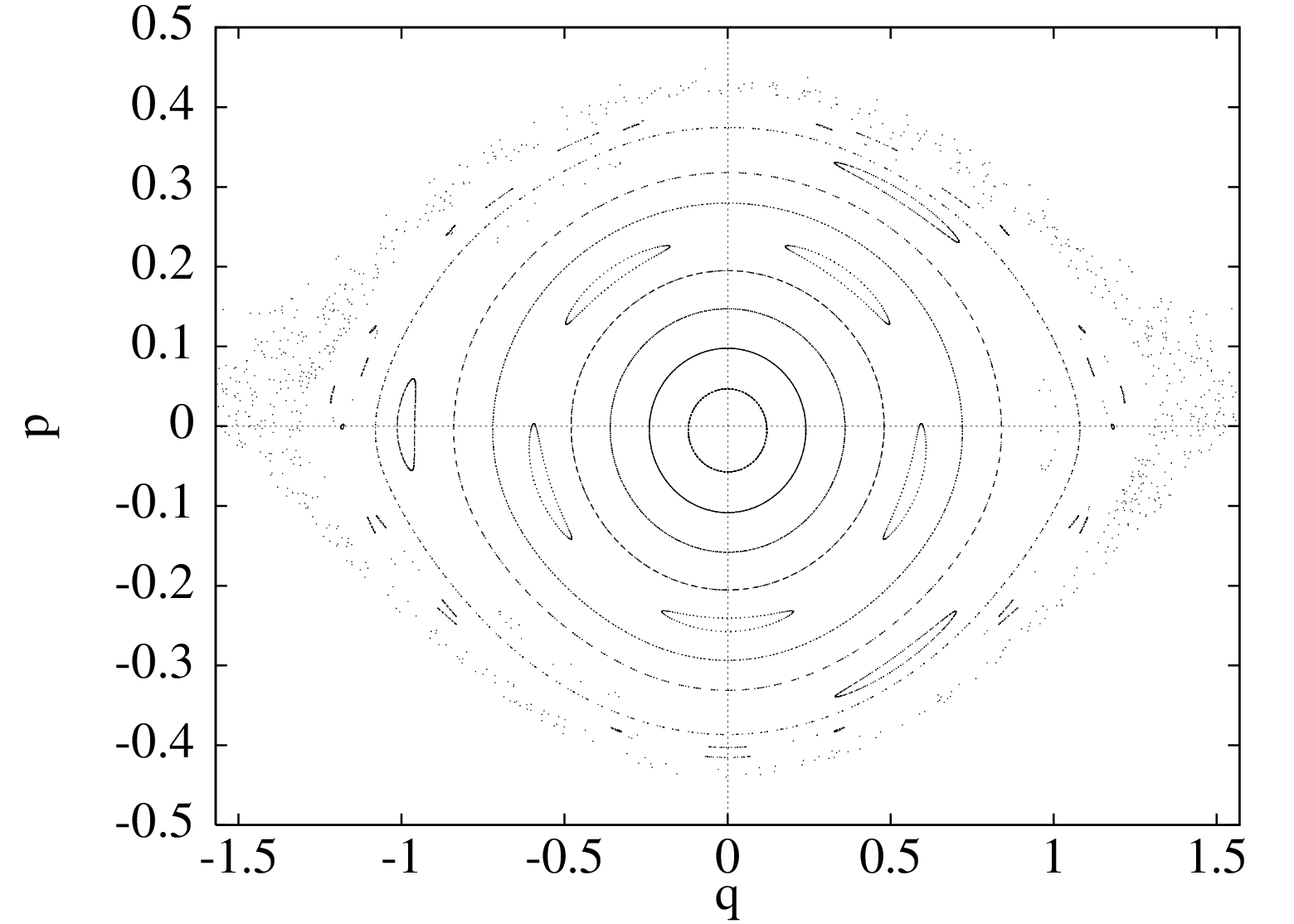,width=3 in}
                 }
      \caption[]{The 
       classical dynamics inside the well. Stroboscopic Poincar\'e map. The 
       values of the parameters are $V_0=0.048$, $\epsilon=0.005$, 
       $\Omega=2$.}
      \label{fig:space}
\end{figure}

The presence of a periodic perturbation in~(\ref{hamiltonian}) breaks 
the integrability of the classical Hamiltonian. The most important 
features of this condition is the appearance of nonlinear resonances 
in the phase space together with regions characterized by extended 
chaotic motion (stochastic layer). The relevance of the chaotic 
motion depends on the strength $\epsilon$ of the perturbation term, 
so that the system can be more or less chaotic. In Fig.~\ref{fig:space}
we show a stroboscopic mapping of the dynamics, namely the position
in phase space at fixed intervals of time which are integer multiples
of the forcing term period $T=2\pi/\Omega$, for a generic weak-chaos case.
Some nonlinear resonances and the 
stochastic layer around the separatrix are clearly visible. 

The nonlinear resonances are the visible consequences of the small 
denominators problem. These are related to the secular terms which 
appear in the perturbative solution of the equation of motion of 
non-integrable systems. In the weak-chaos condition their position
in the phase space can be 
obtained by considering the effect of the time-dependent term as a 
perturbation on the dynamics expressed by the constant Hamiltonian
\begin{equation}
H_{0}=\frac{p^{2}}{2}+V_0 (1-\cos(2q))|_{-\pi<q<\pi} \ . 
\label{eq:h0}
\end{equation}
The KAM theorem~\cite{kam} states that, as long as the
perturbating term can be considered {\em small},
the main part of the phase space remains 
practically unperturbed and that only the tori which are resonant 
with the forcing term are destroyed and replaced by chains of islands 
like the ones shown in Fig.~\ref{fig:space}. The resonant condition 
can be written as 
\begin{equation}
\omega_{0}(E)=\frac{m}{n}\Omega\ ,
\label{eq:reso}
\end{equation}
where $n$ and $m$ are integer numbers and $\omega_{0}(E)$ is the 
frequency of the unperturbed motion inside the well which depends on 
the energy $E$. In our case it is possible to express $\omega_{0}(E)$ 
in term of the elliptic function $K(k)$ as
\begin{eqnarray}
\label{eq:omegae}
\omega_{0}(E)=\pi \sqrt{V_{0}}/K(k) \\ 
k\equiv (E+V_{0})/2V_{0} \nonumber\ .
\end{eqnarray}
The result of Eq.~(\ref{eq:reso}) would actually indicate that all 
the tori are destroyed by the perturbation, being the rational 
numbers dense among the real ones. Nevertheless, the KAM 
theorem
assures that the effects of the perturbation becomes smaller and 
smaller with increasing the order of the resonance, i.e., with 
increasing the numerator $m$.
This is clearly visible in 
Fig.~\ref{fig:space} where we can only 
recognize the chains corresponding to $m=1$, i.e., the $1/5$, $1/6$ 
and $1/7$ resonances.
However, even for small perturbation, in the neighborhood of the 
separatrix of the unperturbed system the motion is always dominated 
by chaotic dynamics. In other words, trajectories which in the 
absence of perturbation are bounded inside the potential well, can 
now overcome the energy barrier and eventually escape from the well. 
In the generic weak chaos
condition, the stochastic layer around the separatrix is dynamically 
separated from the phase space region corresponding to bounded 
trajectories by unbroken tori, so that the process of escape driven 
by chaos is limited in phase space. Therefore, in the classical case, 
the decay of the population of the well is possible only if some 
particles are initially put inside the stochastic region. This 
process have been extensively studied in the last years in various 
classical and quantum models, the most known of which is probably the Hydrogen atom in the 
presence of a strong radiation field\cite{idro,ionization}.
In that system the dynamic process described above leads to the 
ionization of the atom. However, in this paper we want to focus on 
the connection between chaos and processes which would be classically 
impossible, such as quantum tunneling. For that reason we shall 
analyze the decay of the well population for a quantum particle 
initially put in the phase space region which 
corresponds, even in the presence of chaos, to bounded motion. In 
connection to this it is worthwhile to remember that in quantum 
mechanics the situation is never as simple.
Whatever the initial condition is, the wave function cannot be 
sharply located inside a finite region but exhibits smooth decreasing 
tails which extend over the stochastic layer. Therefore, to keep the 
classical chaotic diffusion process as small as possible, we shall 
consider a weak-chaos regime with a small stochastic layer like in 
Fig.~\ref{fig:space}, and in addition to this, we shall study the 
decay of quantum states well localized inside the potential well.

\section{The Model: Quantum Dynamics}
\label{sec:quantum}
We studied the quantum decay from the well of Fig.~\ref{fig:poten} by 
integrating numerically the time dependent Schr\"odinger equation 
associated with Hamiltonian~(\ref{hamiltonian}). This can be done 
using a FFT splitting algorithm~\cite{split} and using absorbing 
boundary conditions~\cite{kosloff} as described, for example, in 
Ref.~\cite{fendrik}.

In order to single out the effect of the chaotic perturbation on the 
process of quantum decay, it is necessary to choose as initial 
condition a state localized inside the well which
has an unperturbed dynamics as simple as possible. 
The eigenstates of Hamiltonian~(\ref{hamiltonian}) with $\epsilon=0$ 
are not useful in this context being system ~(\ref{hamiltonian}) an open  
system (continuous spectrum with stationary states which do not have 
finite support inside the well). We thus resorted to use as initial 
condition metastable states which have a negligible {\em  internal 
dynamics} and a long enough unperturbed life-time inside the well. 
These are the {\em resonances} of the potential well of 
Fig.~\ref{fig:poten} defined in the quantum theory of scattering. 
In addition to this, it is worthwhile to notice that, since in the 
parameter range we explored the unperturbed decay probability is 
negligible, the adoption of the eigenstates of 
Hamiltonian~(\ref{eq:h0}) supplemented by periodical boundary conditions
leads essentially to the same results.  
Therefore, as a first approximation we can consider the initial states 
as stationary state of the unperturbed system. For the sake of 
simplicity in the following we shall refer to them as ``eigenstates'' of 
the unperturbed Hamiltonian.
\begin{figure}
     \centerline{\psfig{figure=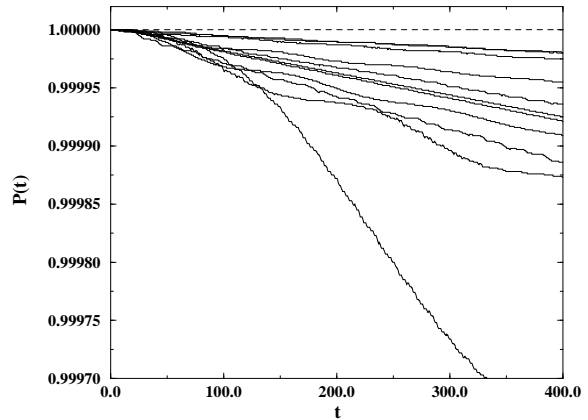,width=3 in}
                }
     \caption[]{The time 
      evolution of the well population. The different curves corresponds to 
      $20$ different values of $\Omega$ included between $\Omega=1.5$ and 
      $\Omega=2.5$. The values of the parameters are $V_0=0.048$, 
      $\epsilon=0.005$, $\hbar=0.025$ and the initial state is the fourth 
      eigenstate of the unperturbed well. The horizontal dashed line 
      correspond to the unperturbed $\epsilon=0$ case. Note the fast 
      oscillations in the population decay which reflect the oscillation of 
      the forcing term}
     \label{fig:decay}
\end{figure}
In Fig.~\ref{fig:decay} we show the time evolution of the population 
inside the well $P(t)=\int_{-\pi/2}^{\pi/2} |\psi(q,t)|^2 dq$ for 
different values of
the forcing frequency $\Omega$. This figures has been obtained by 
choosing as initial condition the fourth eigenstate, $|3\rangle$, of 
the unperturbed well. It can be easily realized that the decay 
probability can be strongly enhanced by the forcing term even in the 
small perturbation regime ($\epsilon=0.005$).
Note that in the unperturbed case the population decay is not visible 
in the scale of this figure, the population remaining practically 
unchanged in the studied interval of time. This shows that the time 
scale of the unperturbed process is much longer than the maximum time 
which we explored numerically.

\begin{figure}
\centerline{
	\vbox{
              \psfig{figure=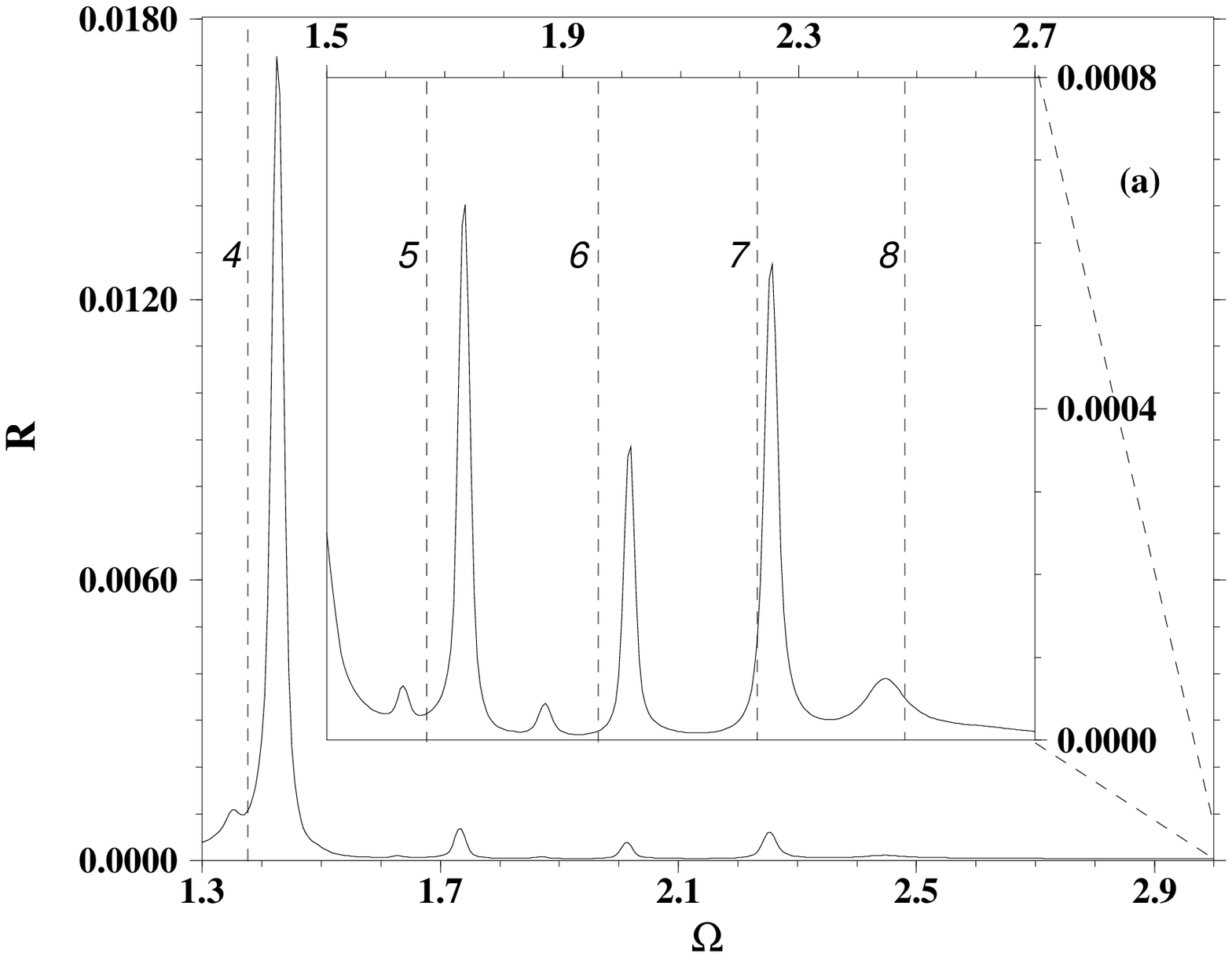,width=3 in}
	      \psfig{figure=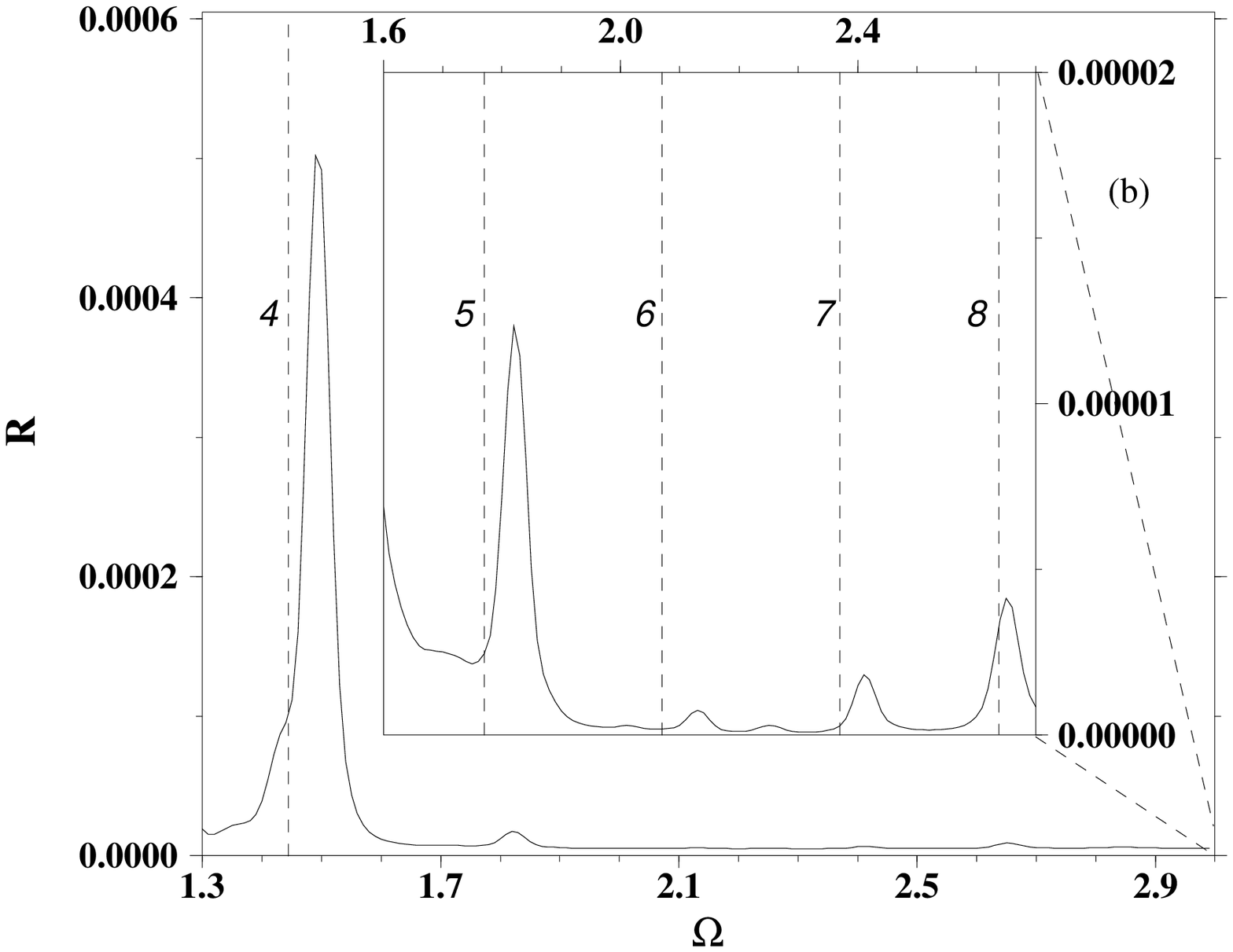,width=3 in}
	     }
	   }
    \caption[]{$R$, calculated with $t_{max}=400$, as function 
     of the driving frequency $\Omega$. The 
     values of the parameters are $V_0=0.048$, $\epsilon=0.005$, 
     $\hbar=0.025$. The two figures correspond to different initial 
     conditions, Fig. a) by choosing as initial state the fourth
     eigenstate of the unperturbed well, Fig. b) by choosing the 
     third. The numbers refer to the classical nonlinear resonances as 
     discussed in Section~\ref{sec:semiclassical}.
     The insets contain enlarged views.}
    \label{fig:pop}
\end{figure}

To obtain a more quantitative representation of the phenomenon we could 
define the decay rate as the inverse of the time integral of $P(t)$, 
namely the inverse of the area contained under the curves of 
Fig.~\ref{fig:decay}. This would imply a very long numerical 
simulation, up to a time where the population has completely leaked 
out. However, we are not interested in the absolute magnitude of the 
decay rate, but only in its relative strength as a function of the 
system parameters. Thus we can simply calculate the time integral of 
$P(t)$ up to a certain time $t_{max}$ and measure the rate 
by studying the quantity
\begin{equation}
\label{eq:rate}
R=1-\frac{1}{t_{max}} \int_0^{t_{max}}P(t) dt
\end{equation}
as a function of the frequency $\Omega$ of the forcing term. As shown in
Fig.~\ref{fig:pop},
$R$ shows a sequence of peaks, similar to a resonant dependence on 
the frequency of the perturbation. We repeated the calculation for 
two different initial conditions, namely by choosing the third and 
the fourth eigenstate of the unperturbed well. The decay rate for the 
third state is smaller, as expected, since this state is more deeply 
located in the potential well, but in both cases we found a similar 
behavior even if the position of the peaks and their intensity is not 
the same.

Except for the presence of the peaks, the figures shows that, as a 
general tendency, the decay increases as the forcing frequency decreases. 
This can be understood using arguments based on the classical dynamics 
of system~(\ref{hamiltonian}). Indeed, as shown in Fig.~\ref{fig:psclass},
the chaotic features of the classical phase space (i.e., width of 
the stochastic layer and of the nonlinear resonances) 
increase by reducing $\Omega$, thus demonstrating that the forcing 
term becomes more important when $\Omega$ gets smaller. Eventually, 
for extremely small $\Omega$, the stochastic layer becomes so wide that, 
for the chosen initial conditions, the escape from the well via 
the direct classical process becomes the dominant process. Since we 
are rather interested in the quantum mechanism of escape from the well, 
we shall not explore the condition of small $\Omega$'s corresponding 
to strong classical chaos.

In addition to this, there is a further reason to limit the analysis 
to $\Omega$'s not too small. Our evaluation of the decay rate,
as the integral of 
the surviving population in the well over a finite time, is meaningful 
only if the time of integration is much longer than the period of the 
perturbation, and therefore we shall limit ourselves to study the 
range of large $\Omega$ $(2\pi/\Omega << t_{max})$.

However, it is important to point out that the numerical results 
represented in Figs.~\ref{fig:decay} and~\ref{fig:pop} does not account 
for the process of classical escape from the well, even to the 
smallest used value of $\Omega$. A direct calculation demonstrates 
indeed that the classical decay is always negligible in the 
considered parameter range of $\Omega$. This can be easily assessed 
by numerically integrating system~(\ref{hamiltonian}) using a classical 
initial particle distribution mimicking the phase-space 
representation of the initial quantum state (see 
Section~\ref{sec:comparison} for details). Even for the smallest value 
of $\Omega$ used to obtain the results of Figs.~\ref{fig:decay} 
and~\ref{fig:pop}, the classical population in the well does not practically
change with time. 

Therefore, the peaks and the general tendency of Figs.~\ref{fig:decay} 
and~\ref{fig:pop} are genuine quantum effects, but, while the latter 
can be associated to the increased effectiveness of the forcing term 
in~(\ref{hamiltonian}) with decreasing $\Omega$, the former effect does 
not have any simple interpretation. The understanding of this will be 
the subject of the next section.

To conclude the 
analysis of Fig.~\ref{fig:pop} we have to assess the role of the 
quantum decay process in the absence of the external perturbation 
(i.e., $\epsilon = 0$). By direct numerical integration we observed 
that in the unperturbed case the decay rate is orders of 
magnitude smaller than that of the forced case. Indeed, for $\epsilon = 
0$, the population inside the well remains equal to one (within the 
numerical precision) even at $t=t_{max}$. This confirmed that within 
the observed time, the chosen initial states can indeed be regarded 
as ``stationary states'' of the unperturbed system. On the other hand, 
this result, as well as previous observations, suggest that the sudden 
increase of the decay rate is connected to the classical chaos in a 
way which is not directly connected to the classical 
process of barrier crossing 
via chaotic diffusion through the stochastic layer. In the next 
section we shall build up a theory to explain the presence of peaks.

\section{A semiclassical analysis}
\label{sec:semiclassical}
The results shown if Fig.~\ref{fig:pop} resemble the typical CAT 
behavior: when a system parameter is changed the decay rate 
presents an irregular sequence of peaks on a smooth 
background. It is thus natural to look for a connection between CAT 
systems and our model. First of all, we notice that, due to the
continuous nature of spectrum, 
the level dynamics of our system can not be simply described in terms 
of avoided crossings. Nevertheless, we believe that the phenomenon 
underlying Fig.~\ref{fig:pop} retains much of the features of the 
CAT processes. In particular we think that the 
argument introduced in Ref.\cite{prenoi} to explain the tunneling
irregularities of a quasi-integrable system, can be effectively used 
also in this model. In the cited reference, a connection between the peaks in 
the tunneling rate and the position of the nonlinear resonances in 
the classical phase space was found.
In this section we shall exploit the same idea.

The analysis of Ref.\cite{prenoi} follows the line of 
Refs.~\cite{semitunnel,breuer} and is based on a semiclassical 
approximation which makes use of simple arguments. We shall assume that 
the classical system is only weakly perturbed by the external 
perturbation: the size of the chaotic region is considered to be 
small with respect to the portion of the classical phase space 
covered with regular trajectories so that the main effect of the 
perturbation is the appearance of chains of nonlinear island in the 
classical phase space as shown if Fig.~\ref{fig:space}.
In this condition, the area of the 
regular region, that is the phase-space region which is encircled by 
the last unbroken torus inside of the stochastic layer, is much larger than 
$\hbar$. In this way, the regular region can accommodate several 
quantum states and the semiclassical approximation becomes 
meaningful.

As we discussed in the Introduction, all the studies about the effects of 
chaos on tunneling or on decay processes were concerned with the 
role of the stochastic layer and the classical transport therein as 
the main contribution to the barrier crossing.  This effect is surely 
present, but its contribution is not always the most important, at 
least in a weak-chaos regime.  In fact as shown in Ref.~\cite{prenoi}, 
by studying the tunneling process in a forced double well system in a 
condition on weak chaos, it has been possible to clarify the role of 
until now disregarded actors: the nonlinear resonances which are 
present inside the regular region of the well.  These 
pre-chaotic structures cannot in fact contribute to the classical 
transport over the barrier, because they are embedded in a regular 
region of unbroken tori, but they can perturb the quantum dynamics via 
a indirect process similar to the one discussed in CAT.  The process 
of avoided crossing responsible of the tunnel irregularity in CAT can 
in fact be present also in a weak-chaos regime, with the difference 
that the third state which, by crossing, modifies the tunneling, 
needs not to be chaotic.

This process, which we think to be a special realization of the more 
generic CAT one, in the weak-chaos regime can give a contribution to 
the tunneling rate modification even more important than the one 
connected to the chaotic region of the phase space.  Moreover, due to 
the fact that the perturbing third state is not chaotic, it is 
possible to obtain a quantitative prediction of the avoided crossings 
and hence on the positions of the tunneling irregularities.  We shall 
now review the derivation of this prediction which has been only 
sketched in~\cite{prenoi}.

Let us note first of all that the Hamiltonian under study is time 
dependent and that this prevents us from adopting the 
Einstein-Brillouin-Keller (EBK) quantization conditions~\cite{ebk} in 
their original form.  However, following the work of Breuer and 
Holthaus~\cite{breuer}, who, in turn, extended the method of the 
canonical operator as developed by Maslov and Fedoriuk~\cite{mp81}, it 
is possible to adapt the EBK prescriptions to periodically 
time-dependent systems.  The generalization of Ref.~\cite{breuer}
is based on the prescription of 
Arnold~\cite{a88} and leads to semiclassical quantization rules for 
the Floquet quasi-energies and quasi-eigenstates~\cite{floq}.
This is made possible by a suitable extension of 
the phase space, including the time $t$ as a coordinate and adding the 
corresponding conjugate momentum.  In the one-dimensional case the 
semiclassical quantization prescriptions read:
\begin{eqnarray}
J &=&\frac{1}{2\pi} \oint_{\gamma 
_{1}}pdq=\hbar (n+\frac{\nu}{4}),
\label{semiclassical} \\
E_{n,m} &=&-\frac{1}{T}\int_{\gamma _{2}}(pdq-Hdt)+\hbar \Omega m.  \nonumber
\end{eqnarray}
The meaning of the symbols adopted in this expression can be explained 
by remembering that in the one-dimensional case the extended space is 
the tridimensional phase space $\{q,p,t\}$.  A regular trajectory is 
contained on a flux tube in this tridimensional space, flux tube which 
repeat itself periodically along the $t$ direction.  Thus $J$ is the 
action associated to the quantized trajectory, $\gamma_{1}$ is a 
close path winding once around the flux tube and lying in the plane at 
a given time $t$ and $\gamma _{2}$ is a path stretching itself 
out on the surface of the flux tube such that it can be continued 
adopting periodic boundary conditions. In other words, the path 
$\gamma _{2}$ in the extended tridimensional phase space, moves from 
an initial point lying on the plane $t=0$ to a final point lying on 
the plane $t=T$.  Note also that we can choose to lay the path $\gamma_{1}$
on the Poincar\'{e} section of the flux tubes.  Finally, if we restrict 
ourselves to consider the closed orbits inside the potential well, the 
Maslov index assumes only the value $\nu=2$.
  
Worth of a detailed discussion is the structure of the quantized 
energy $E_{n,m}$ (for simplicity we shall use the terms {\em energy} and 
{\em state} as equivalents of {\em quasi-energy} and {\em quasi-state}).
While the index $n$, which values are fixed by the 
first of Eqs.~(\ref{semiclassical}), has the usual role of principal 
quantum number, the index $m$, and the dependence of $E_{n,m}$ on this 
one, reflects the periodicity of the time dependent term of the 
Hamiltonian. In fact, an important aspect of the Floquet theory is the 
Brillouin zone structure of the energy spectrum: for each physical
solution labeled by $n$ we have a infinite series of representative 
labeled by the value of $m$. Naturally all the physical information 
is contained in the first Brillouin zone $0 \leq E_{n,m} < \hbar 
\Omega$, or equivalently, we can say that any solution of 
Eqs.~(\ref{semiclassical}) can be folded back to the first Brillouin 
zone by an appropriate choice of $m$.

At this point it is important to recall that the earlier formalism 
represents a valid quantization procedure only in closed systems,
where the notion of energy 
levels is meaningful, and where the phase space is filled with 
regular tori. However, we are investigating the 
properties of states that lie well inside the stability region, and 
that present a small decay probability (see Fig.~\ref{fig:pop}). In 
this condition, we believe that an analysis which, according to the
prescriptions of CAT, connects the tunneling irregularities with the 
presence of avoided level crossings, can retain its validity.
Let us thus proceed disregarding the continuity of the spectrum and
looking for the presence of level 
crossings in the unperturbed hamiltonian spectrum as a function of the 
forcing frequency $\Omega$.

Following Ref.~\cite{breuer} we can look for the level 
crossings by replacing $H$ with $H_{0}$, 
let us in fact recall that we are always considering a perturbative approach.  
The condition of crossing between states $n$ and $n'$ 
in the Brillouin zone yields the following equation:
\begin{equation}
H_{0}(\hbar (n+1/2))+\hbar \Omega m=H_{0}(\hbar (n^{\prime 
}+1/2))+\hbar \Omega m^{\prime }\ . 
\label{degenera}
\end{equation}
This equation can be simplified if we assume that $\hbar$ is so 
small as to make negligible the quantity $\hbar (n-n^{\prime })$. If 
we now expand the RHS of this equation around this small parameter, we 
obtain
\begin{equation}
\frac{dH_{0}}{dn}(n-n^{\prime })+\hbar \Omega m=\hbar \Omega 
m^{\prime }
\label{resona1}
\end{equation}
which can be rewritten as
\begin{equation}
\frac{dH_{0}}{dJ}\frac{dJ}{dn}(n-n^{\prime })=\hbar \Omega (m^{\prime 
}-m)
\label{resona2}
\end{equation}
or, by using $J=\hbar (n+1/2)$,
\begin{equation}
\omega_{0}=\frac{\Delta m}{\Delta n}\Omega +{\cal O}(\hbar )\ ,
\label{resonance}
\end{equation}
where $\omega_{0}\equiv \omega_{0}(J)=\frac{dH_{0}}{dJ}$ is the 
frequency of the unperturbed motion as a function of the classical 
action $J$, $\Delta n \equiv n^{\prime }-n$ and
$\Delta m \equiv m-m^{\prime }$.  
The condition of levels crossing, in the limit of vanishing $\hbar$ 
can thus be obtained by solving the classical equation which
corresponds to the condition for the onset of the classical nonlinear resonances.

This is a very important results, because it means that, for 
sufficiently small $\hbar ^{\prime }s$, there is a correspondence 
between the presence of a nonlinear resonance in the classical phase 
space and a level crossing in the spectrum and therefore a 
correspondence between the tunneling irregularities and the nonlinear 
resonances.  In other words, the tunneling peaks are the quantum 
manifestation of the non-integrable behavior of the underlying 
classical dynamics.  The results of Eq.~(\ref{resonance}) is even more 
specific, in fact is says that the crossings of the two unperturbed 
levels $E_{n,m}^{0}$ and $E_{n^{\prime },m^{\prime }}^{0}$ is related 
to the superposition between the semiclassical quantization torus of 
one of the two states and the nonlinear resonance of appropriate order 
$\Delta m/\Delta n$.

This process admits a simple intuitive representation. Let for 
example consider a quantum state located deep in the well. Its quantization 
torus lies inside the well and the wavefunction in the semiclassical 
regime is mainly located around this torus. The decay rate in the unperturbed 
case can be obtained by the usual semiclassical calculation of the 
probability of barrier crossing. If we turn the perturbation on in a 
regime of weak chaos we have a high probability that nothing happens, 
due to the fact that most of the phase space inside the well remains 
unperturbed. But if we change an external parameter as, for example, the 
forcing frequency $\Omega$, we obtain that the nonlinear resonances 
move in the phase space and, for particular values of the parameter, 
one of them can intersect the quantization torus which is destroyed. 
This would be probably reflected in a perturbation of the quantum 
state and thus in a modification of its decay rate. This is 
exactly what is described by Eq.~(\ref{resonance})

Notice anyway that 
Eq.~(\ref{resonance}) involves resonances of any order and this 
implies that the change of a system parameter makes a chosen level 
undergo a virtually infinite number of crossings.  In other words, 
when we turn the perturbation on, the mathematical curve corresponding 
to a particular energy level becomes extremely complex, 
being fragmented by the avoided-crossing effects in an infinite number 
of points.  However, the perturbation produced by the avoided 
crossings strongly depends on the order $\Delta m$ of the resonance 
and in practice, if we restrict ourself to the perturbative regime, 
the only significant crossings are those associated with the 
first-order resonances ($\Delta m=1$). On the other hand the same 
happens in the classical dynamics, as discussed in 
Section~(\ref{sec:classical}).

The simple prediction of Eq.~(\ref{resonance}) is valid in the strong 
semiclassical limit and is therefore unavailable for a numerical 
check, being the smallest value of $\hbar$ dictated by the computer 
limitations.  The relation between nonlinear resonances and tunneling 
peaks has been indeed proven in Ref.~\cite{prenoi} by means of a 
further expansion of Eq.~(\ref{degenera}).  To do that we proceed by 
evaluating the second order term in $\hbar(n-n^{\prime })$.  We obtain:
\begin{equation}
\omega_0+\frac{1}{2}\hbar \frac{d \omega_{0}}{dJ}\Delta n=
\frac{\Delta m}{\Delta n}\Omega\ +{\cal O}(\hbar^2)\ . 
\label{resocor1} 
\end{equation}
On the other hand,
\begin{equation}
\frac{d\omega_0}{dJ}\equiv \frac{d \omega_{0}}{dE}\frac{dE}{dJ}
=\frac{d\omega_0}{dE}\omega_0\ . 
\end{equation}
Thus we can write~(\ref{resocor1}) as follows: 
\begin{equation}
\omega_0(E)=\frac{\Delta m}{\Delta n}\frac{\Omega}{(1+\frac{\hbar}{2}
\frac{d\omega_0(E)}{dE}\Delta n)}+{\cal O}(\hbar^2)\ . 
\label{eq:reso2}
\end{equation}
Where we can give an analytical expression to 
$\frac{d\omega_0(E)}{dE}$, by using Eq.~(\ref{eq:omegae}), as
\begin{equation}
\frac{d\omega_0}{dE}=\frac{\pi }{4k^{2}\sqrt{V_{0}}} \frac{1}{K(k)}
(1-\frac{E(k)}{k^{\prime }{}^{2}K(k)})\ , 
\label{domegade} 
\end{equation}
where $k^{\prime }{}^{2}=1-k^{2}$. Eq.~(\ref{eq:reso2}) is the main 
results of Ref.~\cite{prenoi} and represents a generalization of the 
classical nonlinear resonance condition, where the frequency ratio is 
renormalized by means of a quantum correction proportional to $\hbar$.

\begin{figure}
\centerline{\psfig{figure=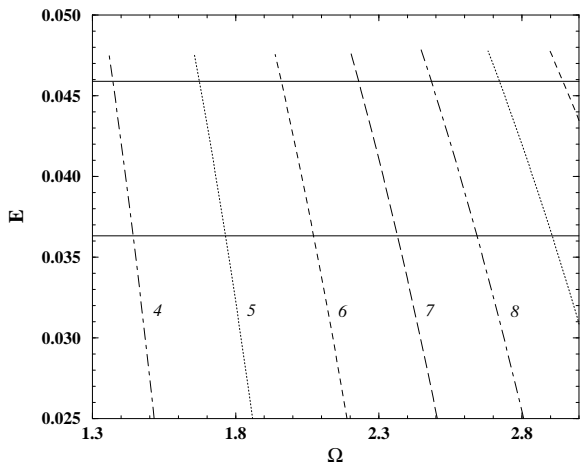,width=3 in}
           } 
\caption[]{The 
solutions of Eq.~(\ref{eq:reso2}) represented by the crossings 
between the solid horizontal lines, indicating the energy of the 
third and fourth eigenstates, and the energy of various nonlinear 
resonances indicated by their order $n$ (dashed and dotted lines). 
The crossings give the theoretical prediction on the position of the 
peaks of decay and are reported in Fig.~\ref{fig:pop} as the vertical 
dashed lines.
The values of the parameter are $V_0=0.048$, $\epsilon=0.005$, 
$\Omega=2$, $\hbar=0.025$.}
\label{fig:cross}
\end{figure}

By using this prediction we are now able to verify our conjecture 
about the validity of this method also in the present case.  From 
Eq.~(\ref{eq:reso2}) it is possible to predict the position of the 
peaks of decay.  After fixing the values of $\Delta m$ 
Eq.~(\ref{eq:reso2}) can be solved as a function of the energy $E$ for 
several values of $n$.  The graphical solution for the first order, 
$\Delta m=1$, crossings is shown in Fig.~\ref{fig:cross} where the 
thick solid horizontal lines correspond to the semiclassical energies 
of the third and fourth eigenstates of the unperturbed 
Hamiltonian~(\ref{eq:h0}), i.e., of the states which we chose as 
initial condition in order to obtain the results of 
Section~\ref{sec:quantum}.  The dotted and dashed curves represent the 
quantum renormalized energies of the classical nonlinear resonances of 
different order $n$ (the order is indicated by the numbers in the 
figure).  The crossings between the horizontal lines and the 
decreasing curves thus indicate the solutions of Eq.~(\ref{eq:reso2}).  
Their position should also indicate the position of the peaks of the 
decay rate.  This is in fact approximately true, as one can check by 
going back to Fig.~\ref{fig:pop}, where we indicated the solutions 
showed in Fig.~\ref{fig:cross} with the numbered vertical dashed 
lines.

We are now able to justify {\em a posteriori} the use of the 
semiclassical theory of this section in the present case. The main 
difference between the physical system of Ref.~\cite{prenoi} and 
Hamiltonian (\ref{hamiltonian}) is the energy levels discreteness. In 
the present work we cannot speak about level crossings and thus the 
calculations above could look invalidated. On the other hand we 
showed that the avoided crossings are nothing more than the {\em 
trait d'union} between the nonlinear classical resonances and the 
peaks of the decay rate:
the presence of the nonlinear resonance produces a level 
crossing which is reflected in a rate irregularity. The same happens
for the system of Hamiltonian~(\ref{hamiltonian}): 
the classical phase space structures 
and the decay rate modifications are related even if the 
intermediate step is less clear due to the continuity of the 
quantum spectrum. Probably we could find a process similar to the 
avoided crossings, but we do not need to look for it as we showed that 
the quantum-classical connection works.

The role of the nonlinear resonances in the perturbation of 
tunneling seems thus to be established also in a system with a continuous 
spectrum, even if the prediction looks approximate.
The not perfect agreement between theory and numerical 
calculation can be traced back to two major approximations. The first 
is the finiteness of $\hbar$ which makes Eq.~(\ref{eq:reso2}) slightly 
inaccurate, while the second and the most important would be the 
approximation related to consider the unperturbed states in 
Eq.~(\ref{degenera}). This last is in fact a double approximation, 
because it disregards the effect of the perturbation, but this is not 
so important as shown in Ref.~\cite{prenoi}, and the effect of the 
unperturbed decay which actually destroys the discrete levels 
picture we used. Nonetheless we think that our results are quite 
clear and to better show the relation between the peaks in the decay 
and the presence of nonlinear resonances disturbing the initial state 
we shall now resort to a graphical picture.

\section{A phase-space representation: quantum-classical comparison}
\label{sec:comparison}
As we said before, in order to connect the quantum dynamics to the 
classical phase-space characteristics we must extend the concept of 
phase space to the quantum case.  This can be done by using a 
phase-space representation of the quantum state and among all the 
different possibilities we chose to use the Husimi representation, 
defined as: 
\begin{equation}
\label{eq:husimi}
\rho(q,p)=\|\int_{-\infty}^{\infty}dx \alpha_{q,p}(x)\psi(x) \|^2 , 
\end{equation}
where $|\alpha_{q,p}\rangle$ is a minimum indetermination state 
(coherent state) centered in $(q,p)$.  Using Eq.~(\ref{eq:husimi}) we 
can obtain a phase representation of a quantum state in terms of the 
positive definite distribution $\rho(q,p)$.  In particular we can 
calculate the Husimi distribution of the initial state which we chose 
to be an eigenstate of the unperturbed well.  This choice will allow 
us to obtain the correspondence we are looking for in a easy way, in 
fact in our approximation, namely, for small decay probability, we can 
safely disregard the dynamics inside the well for the times we 
explored.  This means that the phase-space representation of the 
quantum state practically does not change during the time interval 
considered in Fig.~\ref{fig:pop}, and that in the comparison 
between quantum and classical phase space we can limit ourselves to 
deal with the initial quantum distribution.  

\begin{figure}
\centerline{
\vbox{  \hbox{
        \psfig{figure=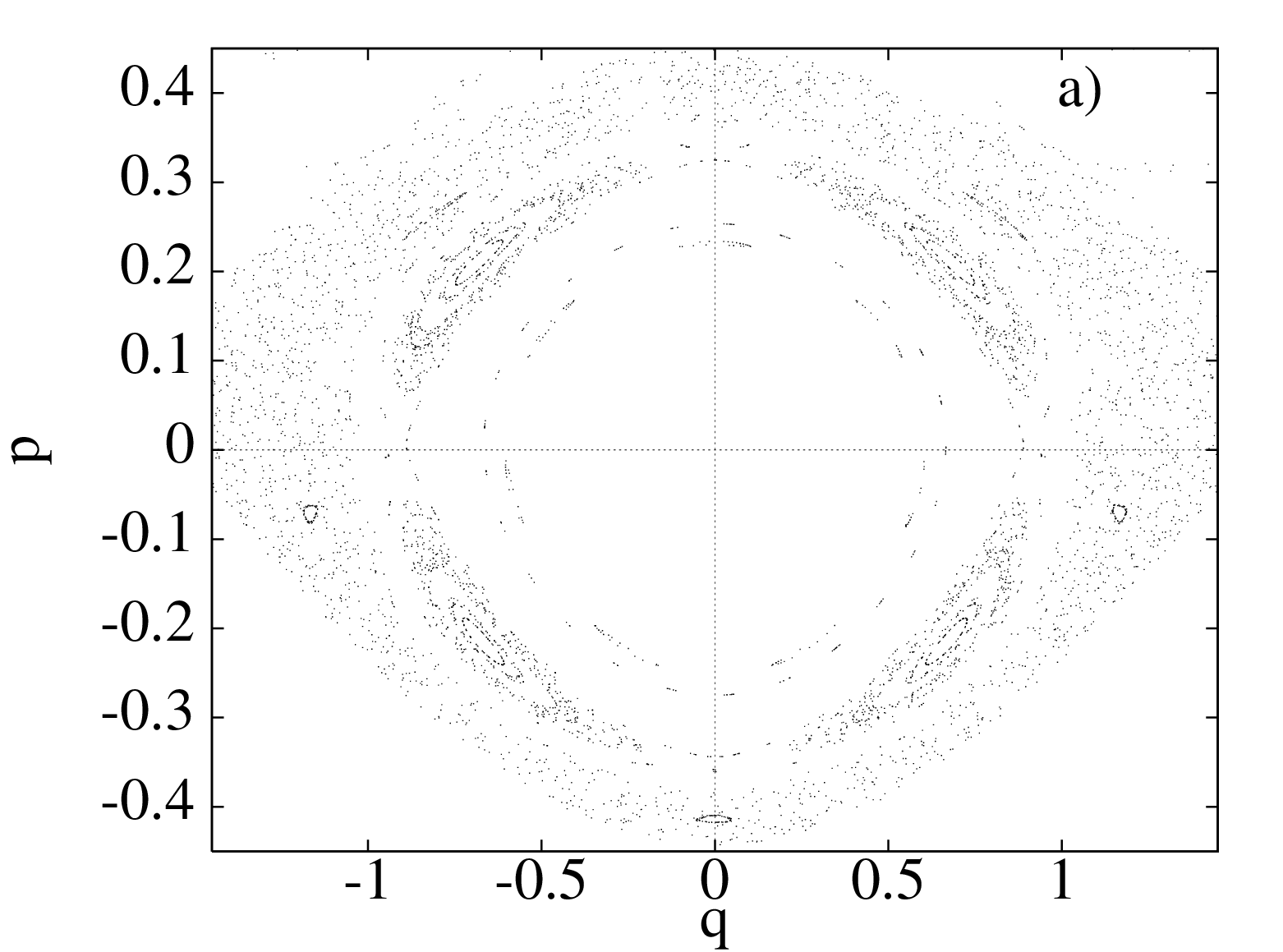,width=1.8in}
        \psfig{figure=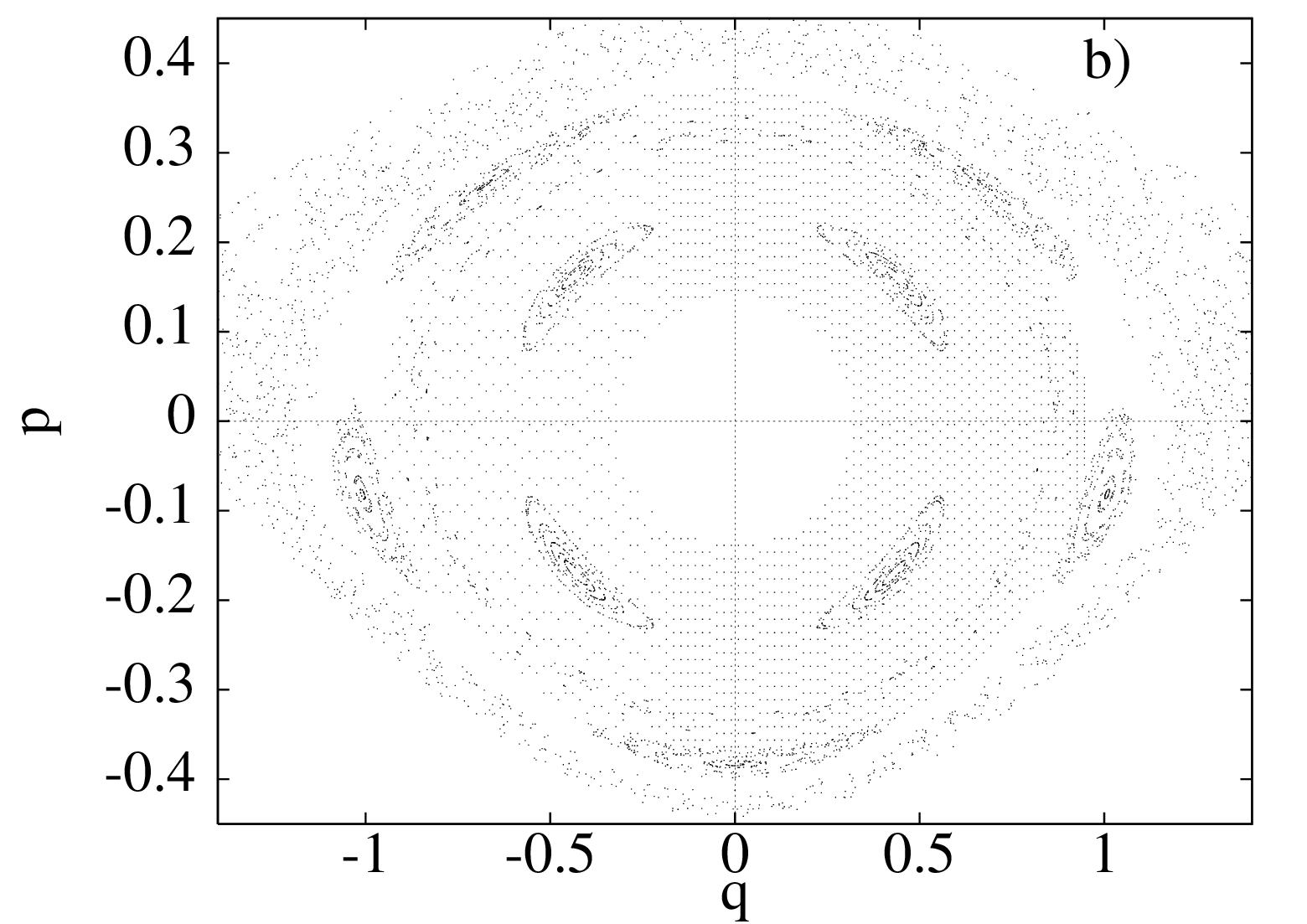,width=1.8in}
        }
        \hbox{
	\psfig{figure=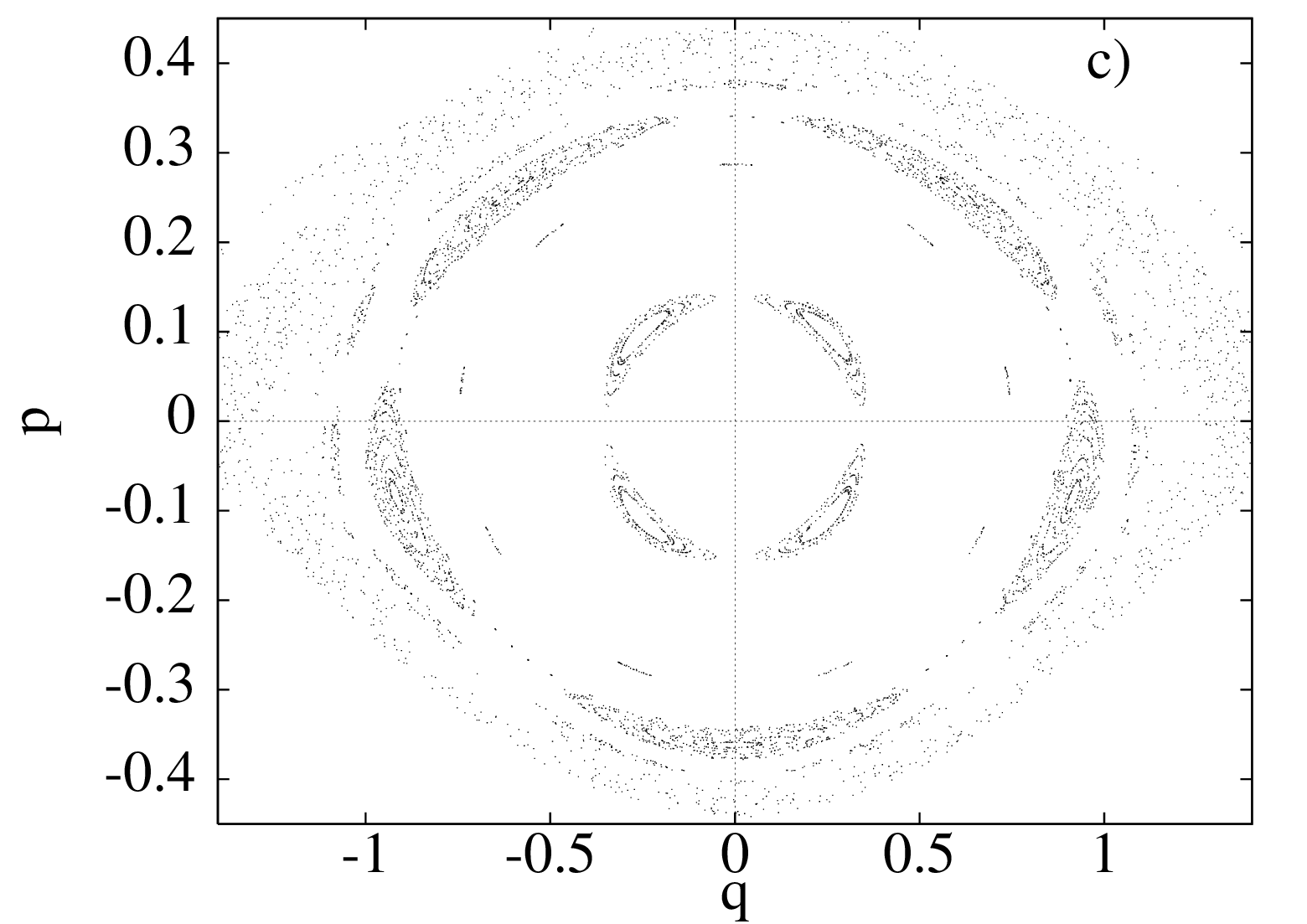,width=1.8in}
	\psfig{figure=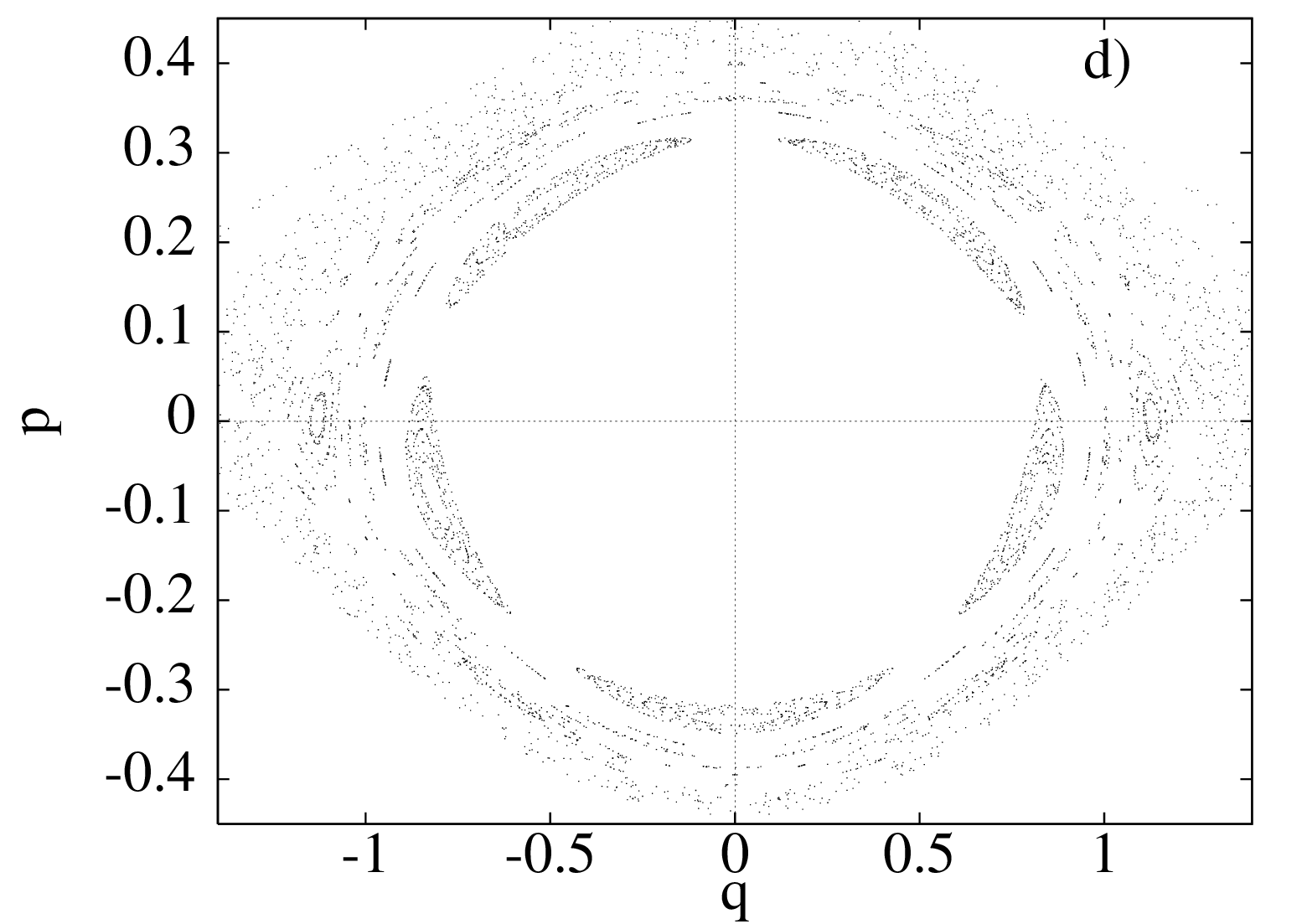,width=1.8in}
	}
	\hbox{
	\psfig{figure=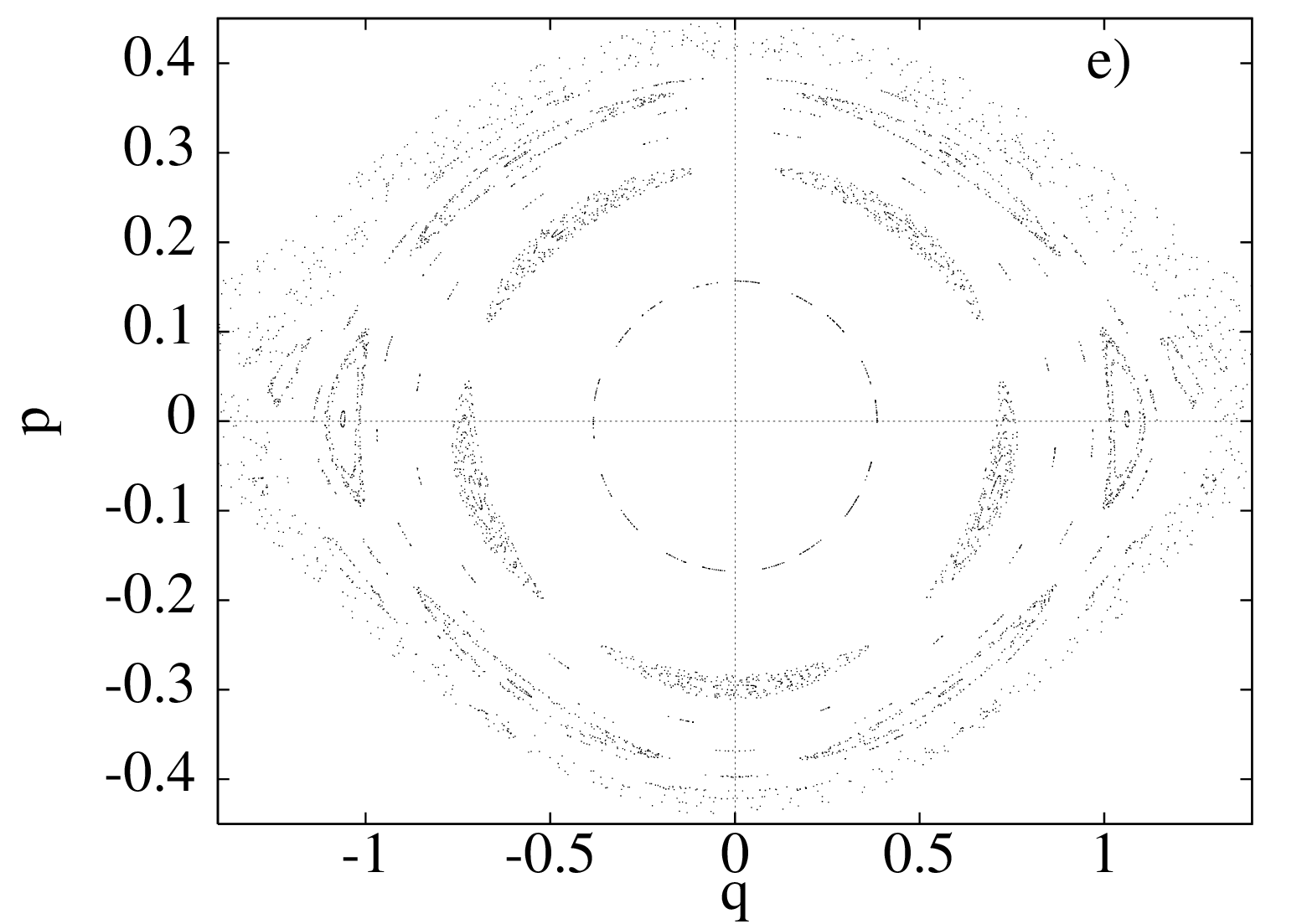,width=1.8in}
	\psfig{figure=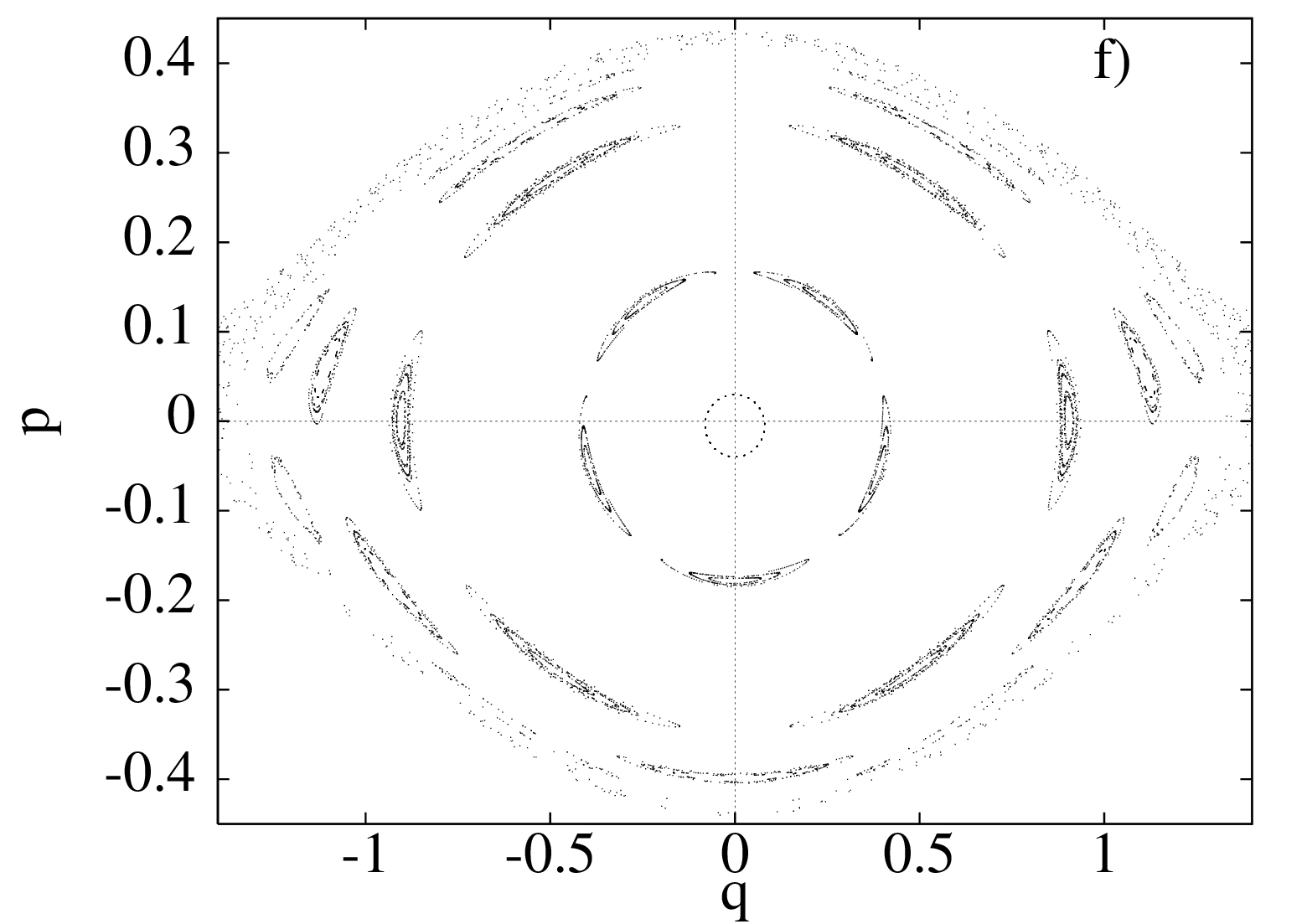,width=1.8in}
	}
	\hbox{
	\psfig{figure=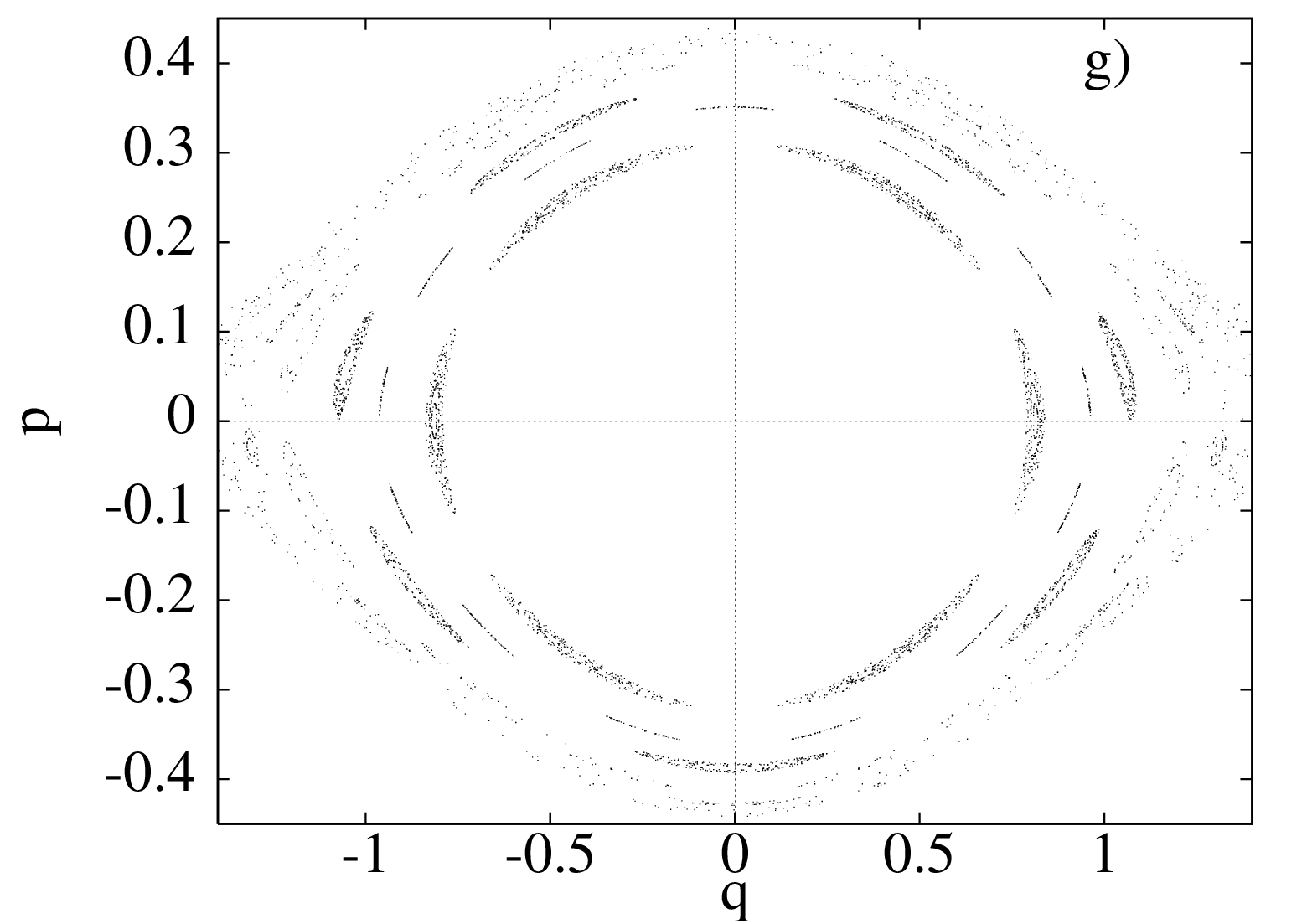,width=1.8in}
	\psfig{figure=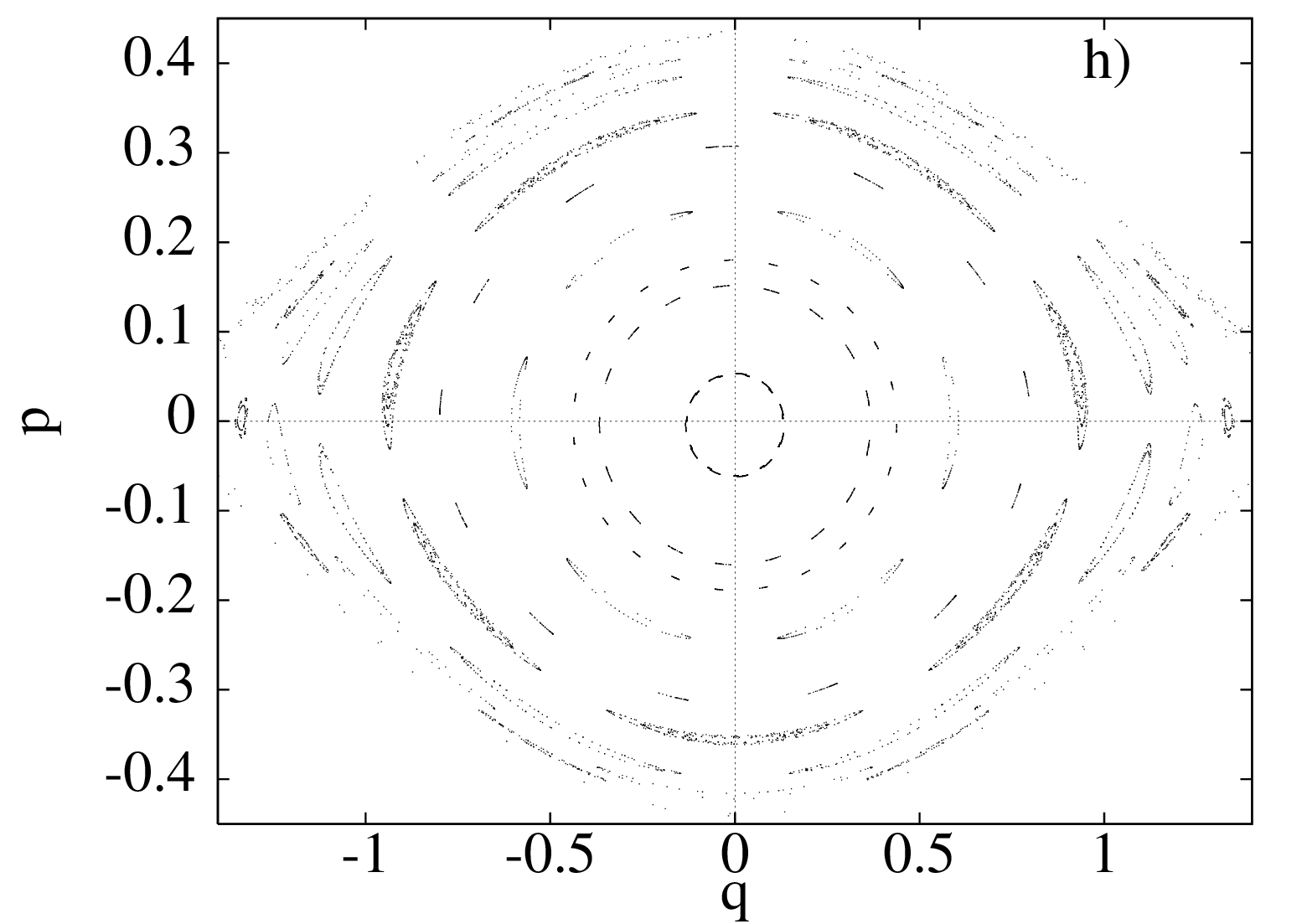,width=1.8in}
	}
	\hbox{
	\psfig{figure=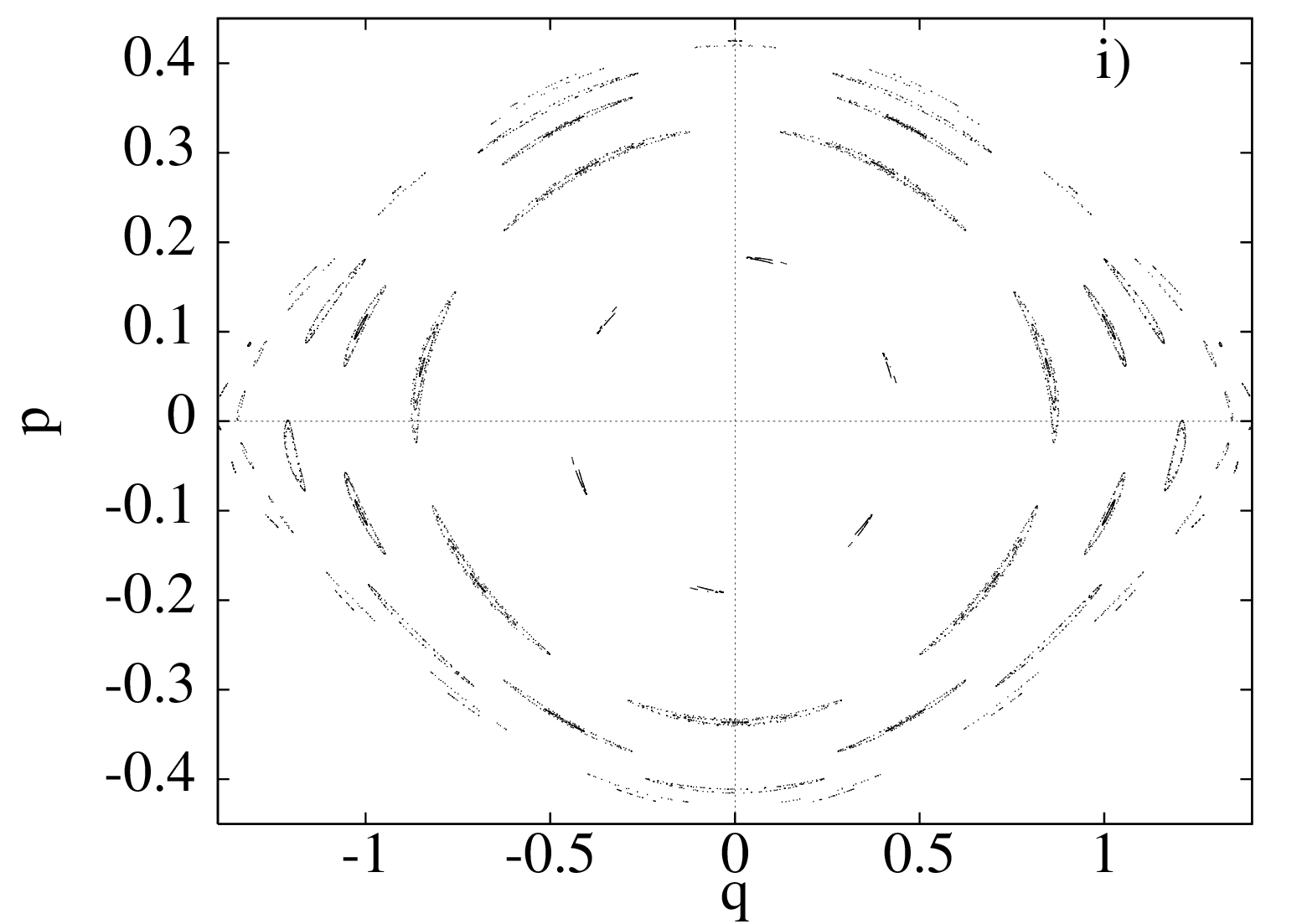,width=1.8in}
	\psfig{figure=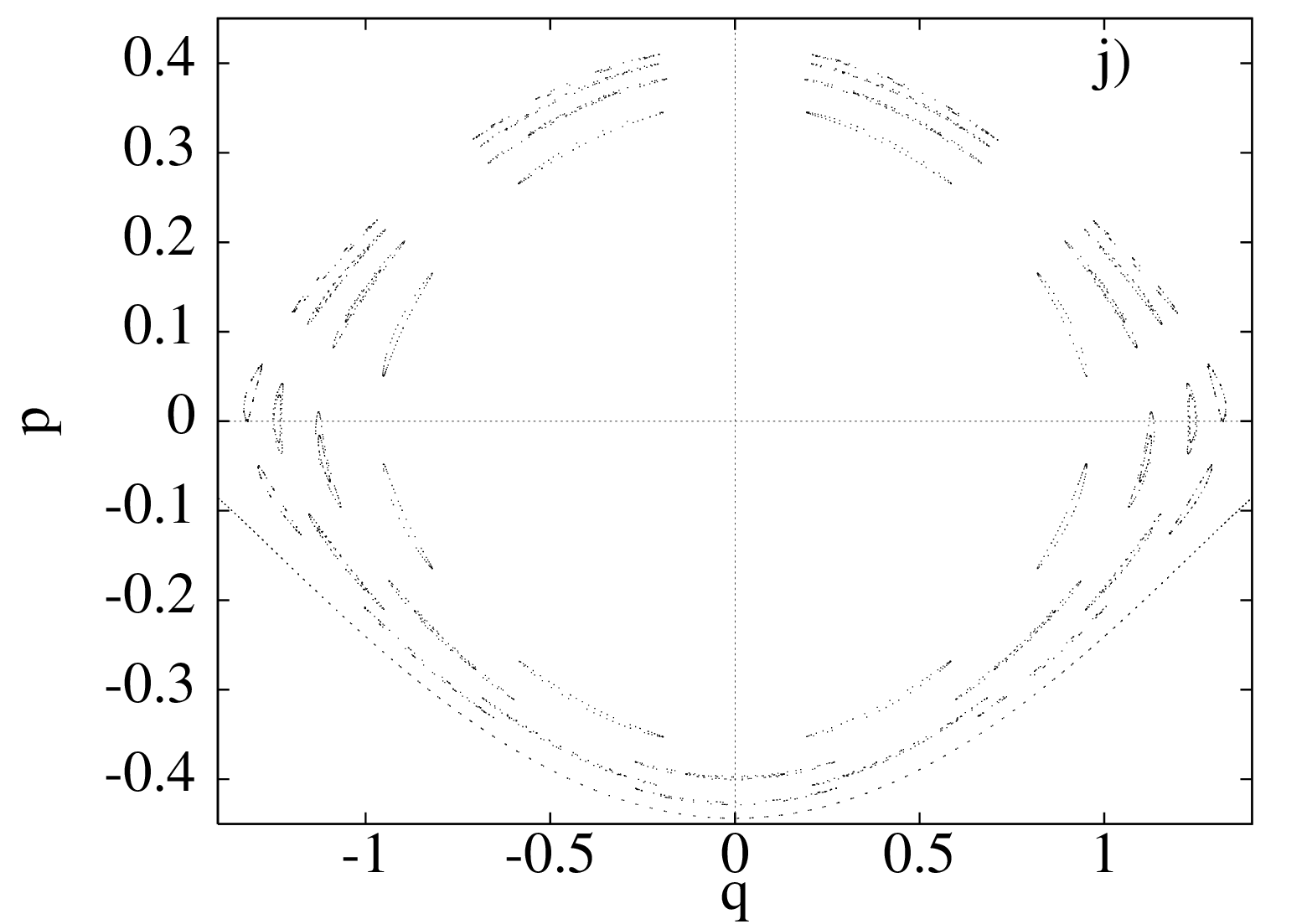,width=1.8in}
        }
       \hbox{}
 }}
\caption[]{Classical phase-space portraits for different values of the 
driving frequency $\Omega$. The values of the parameter are 
$V_0=0.048$, $\epsilon=0.005$, $\hbar=0.025$. The value of $\Omega$, 
increasing from a) to j), is $\Omega=1.4,\ 1.6,\ 1.7,\ 1.8,\ 1.9,\ 
2.1,\ 2.2,\ 2.4,\ 2.5,\ 2.7$, The shaded area in Fig.~b) 
indicates the area under the Husimi distribution of the initial 
state, which in this case is the fourth eigenstate of the unperturbed 
well.}
\label{fig:psclass}
\end{figure}

In Fig.~\ref{fig:psclass} 
we show portraits of the classical phase space for increasing values 
of $\Omega$.  For clarity we draw only the stochastic web and the 
nonlinear resonances island structures, all the rest of the 
phase space being filled with regular tori.  As explained in 
Section~\ref{sec:classical}, due to the choice of the particular form 
of the time dependent perturbation, to the first order in $\epsilon$ 
we have only resonances of the form $\omega_{0}(E)=\Omega/n$ where 
$\omega_{0}(E)$ is the frequency of the motion inside the unperturbed 
well. The frequency $\omega_{0}(E)$ is a decreasing function of the 
energy $E$ for $0 < E < 2V_0$, being equal to $\sqrt{4 V_0}$ for $E=0$ 
and vanishing for $E=2V_0$ which corresponds to the separatrix motion, 
see Eqs.~(\ref{eq:omegae}). This means that as we approach the separatrix
we find resonances of larger order $n$. We can also easily realize 
that as the value of $\Omega$ is increased the nonlinear resonances 
move inside the phase space getting closer to the center of the well, 
and eventually disappearing when the relation~(\ref{eq:reso}) cannot 
be fulfilled any more.
In the meanwhile new resonances appear, moving out 
from the stochastic layer which is the region of the overlapping of 
all the infinite resonances of higher order $n$.

In this motion towards the center of the well, the various resonances 
cross the region of the phase space which is occupied by the Husimi 
distribution of the initial quantum state (shaded area in 
Fig.~\ref{fig:psclass}b) and, as discussed in 
Section~\ref{sec:semiclassical}, we expect this to be related to the 
peaks of Fig.~\ref{fig:pop}. The region of parameters explored in 
Figs.~\ref{fig:pop} and~\ref{fig:psclass} is the same and from a 
first inspection we actually realize that the number of resonances 
crossing the shaded area corresponds to the number of peaks in the 
decay rate. This first result is already convincing but we can 
be more precise by singling out the phase-space portraits which 
correspond to the decay peaks.

\begin{figure}
\centerline{
\vbox{\hbox{
	\psfig{figure=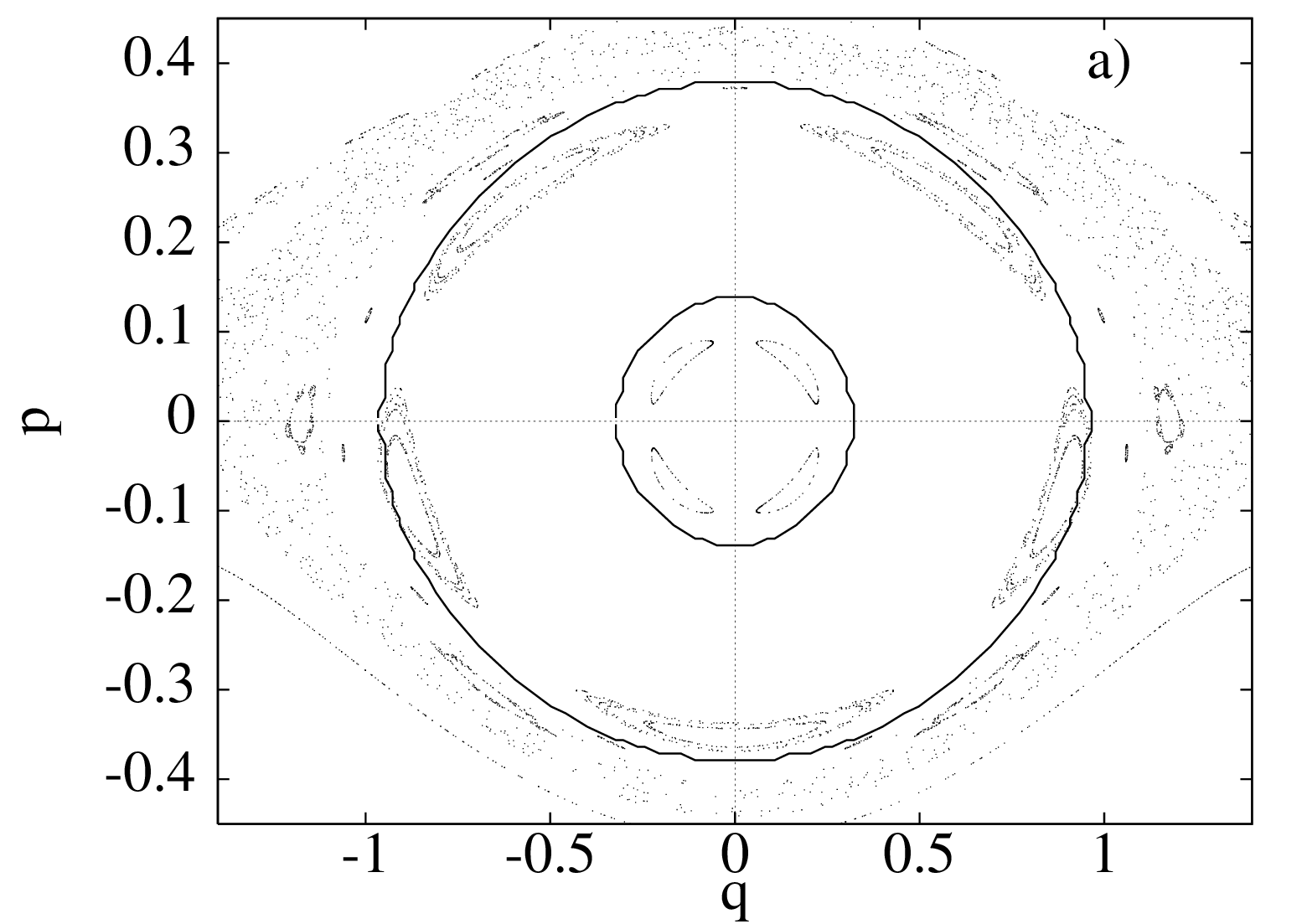,width=1.8in}
	\psfig{figure=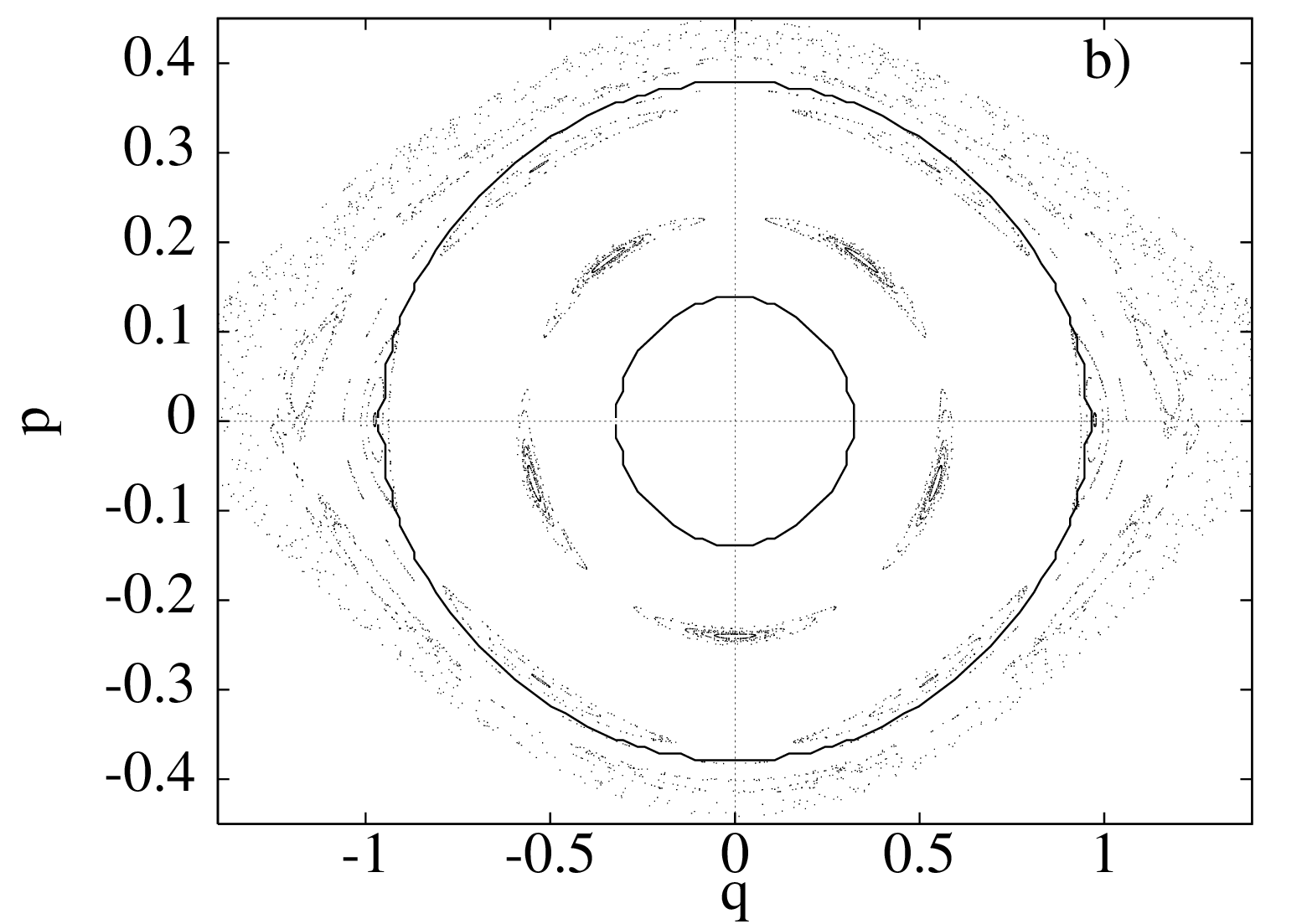,width=1.8in}
	}
	\hbox{
	\psfig{figure=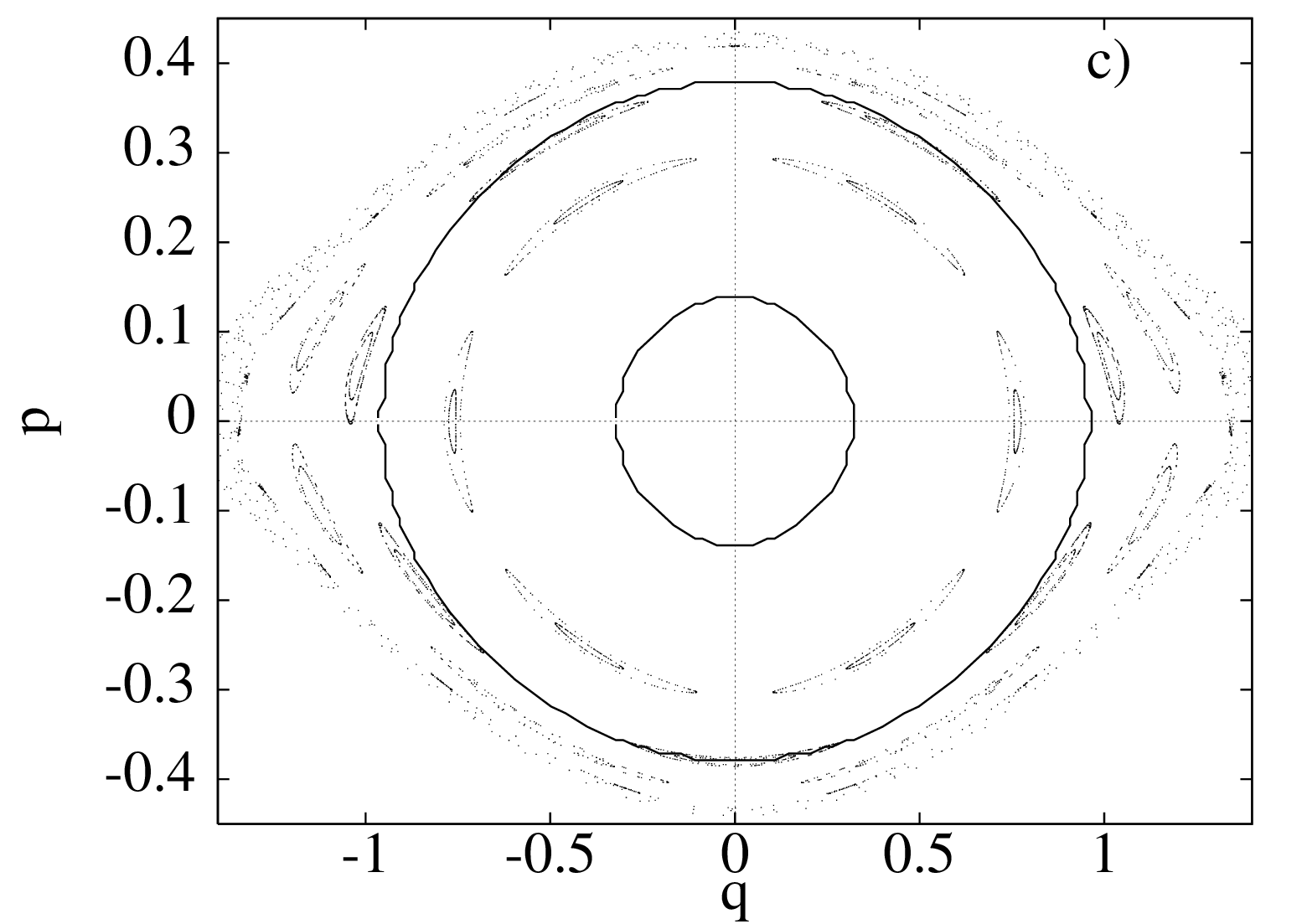,width=1.8in}
	\psfig{figure=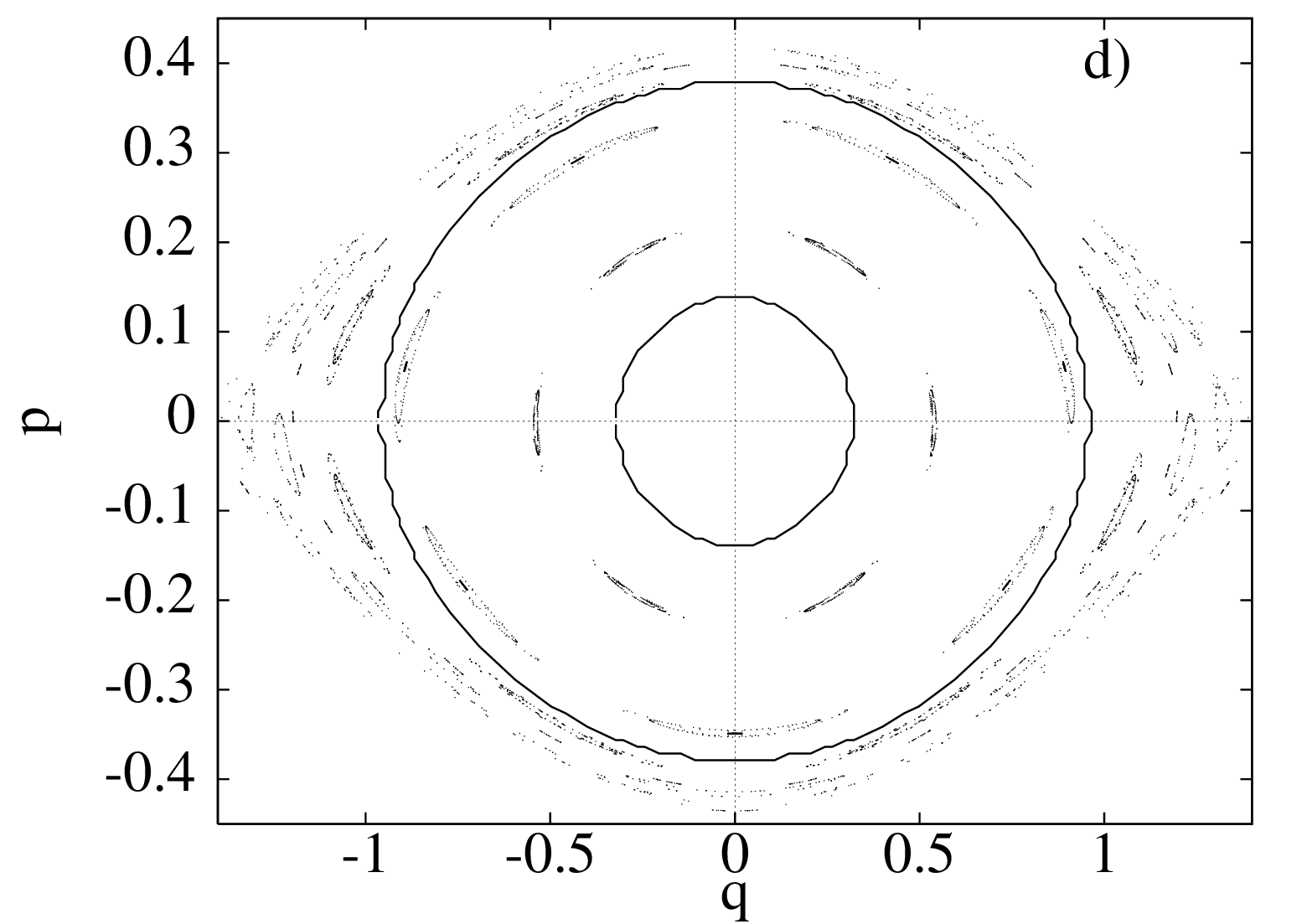,width=1.8in}
	}
        \hbox{
	}
	}}
\caption[]{Classical phase-space portraits for different values of the 
driving frequency $\Omega$. The values of the parameter are 
$V_0=0.048$, $\epsilon=0.005$, $\hbar=0.025$. The values of $\Omega$ 
are chosen as the values corresponding to the peaks in 
Fig.~\ref{fig:pop}a, i.e., $\Omega=1.73\mbox{(a), }2.015\mbox{(b), 
}2.25\mbox{(c), }2.44\mbox{(d)}$. The two continuous circular 
curves indicate the outer and inner border of the Husimi distribution 
of the initial quantum state (the fourth eigenstate).
The border is defined arbitrarily as 
the contour level of the distribution at 0.2 times
the maximum height.}
\label{fig:contour}
\end{figure}

This is done in 
Fig.~\ref{fig:contour} where we show the classical phase-space 
structures and we indicate by the two continuos lines the borders of
the Husimi distribution of the initial quantum state. 
The expected results is confirmed, the peaks in the quantum 
decay rate are related to the modification of the classical phase 
space under the quantum initial distribution by 
a nonlinear resonance. The perturbation seems to be 
effective when a nonlinear resonance enter the 
external border of the initial state distribution and not when the 
nonlinear resonance passes over the center of the distribution where 
the quantization torus is located. This is qualitatively in agreement with 
Eq.~(\ref{eq:reso2}) which predicts that the perturbation appears when the 
nonlinear resonance is close to the semiclassical quantization torus, 
how much close being dictated by the quantum correction which modifies 
Eq.~(\ref{eq:reso2}) with respect to Eq.~(\ref{eq:reso}).  
In fact the quantum correction to the forcing frequency 
$\Omega$ is positive (the term $\frac{d\omega_0(E)}{dE}$
is negative for energy smaller than the barrier heights and thus
the renormalized frequency is always larger than $\Omega$)
and this means that the decay peaks 
should appear for values of $\Omega$ smaller than the values for which the 
nonlinear resonances superpose to the quantization torus. This is 
precisely what happen: the peaks correspond to the approaching
of the nonlinear resonances from the outside.

This result comes along with a 
simple intuitive explanation: when a nonlinear resonance enters the 
external border of the initial state part of the initial distribution 
is moved outward by the islands structure, this produce a higher 
probability of tunneling across the barrier and thus an increase of 
the decay. This effect becomes less important once the nonlinear 
resonance penetrates inside the shaded region; until 
it disappears when the resonance is completely embedded in the 
central part of the distribution.

The results of Section~\ref{sec:semiclassical} have been thus 
confirmed by the comparison of this section, making the picture of 
the role of the classical nonlinear resonances clearer and clearer.

\section{Chaos Assisted Decay versus Chaos Assisted Tunneling}
When we reviewed the numerical results we noticed that we found only 
enhancement of the decay compared to the unperturbed condition.  
This is not the case in the CAT, where the avoided crossings can 
produce both enhancement and decrease of tunneling, which can also 
vanish for particular values of the parameter~\cite{hanggi}. For this 
reason we wrote in the Introduction that the term {\em assisted} used 
in CAT is not really appropriate, but it seems that it could be better 
used in the present contest. To explain this different behavior we 
conjecture that this can be seen as a consequence of the continuity of 
the spectrum in our system. In fact the quenching of the tunneling is 
produced by the accidental degeneracy of the levels of the tunneling 
doublet, as seen in Fig.~\ref{fig:split}. This degeneracy is due to 
the modification induced by the crossing with the third level. In our 
case we do not have discrete levels, but a continuous density of 
states and thus the former picture simply does not apply.  In other 
word for every value of the parameter the modification of the spectrum 
due to the presence of avoided crossings cannot lead to a complete 
degeneracy of the states and thus to a complete quenching of the 
decay. A similar situation and a graphical representation of 
this process can be found in a recent paper~\cite{geiselband}.
On the other hand this continuous spectrum 
characteristic cannot completely rule out the possibility that chaos 
could produce also a decrease of the unperturbed decay rate, which 
we think could be present in some region of the parameters.

\section{Conclusions}
A numerical calculation of the decay probability due to tunneling from 
a potential well showed that in presence of classical chaos the decay 
can be strongly enhanced and that this enhancement depends on the 
system parameters in a resonant-like way. A qualitative inspection of 
the classical phase-space structure revealed a connection between the 
peaks in the decay probability and the presence of classical nonlinear 
resonances in the region of the phase space occupied by the Husimi 
distribution of the initial state. This correspondence has been 
quantitatively explained using a semiclassical result which has been 
shown to be valid in the case of Chaos Assisted Tunneling. We can 
thus conclude that the enhancement of decay from a driven well can be 
explained by means of a perturbation of the region of the phase space 
occupied by the initial state, perturbation produced by the emergence 
of nonlinear resonances. This is a direct connection between the 
modification of a purely quantum effect, the tunneling, and the 
classical phenomenon of destruction of the integrable dynamics which 
is at the basis of the chaotic behavior.


\begin{references}
\bibitem{dh81} D.J. Davis and E.J. Heller, J. Chem. 
Phys. {\bf 74}, 246 (1981).
\bibitem{lb90} W. A. Lin and L. E. Ballentine, Phys. Rev. Lett. {\bf 
65}, 2927 (1990).
\bibitem{hanggi} F. Grossmann, P. Jung, T. Dittrich, and P. 
H\"{a}nggi, Z. Phys. B, {\bf 84}, 315 (1991); F. Grossmann, T. 
Dittrich. P. Yung, and P. H\"{a}nggi, Phys. Rev. Lett. {\bf 67}, 516 
(1991). 
\bibitem{btu93} O. Bohigas, S. Tomsovic and D. Ullmo, Phys. 
Rep. {\bf 223}, 43 (1993).
\bibitem{tu94} S. Tomsovic, D. Ullmo, Phys. Rev. E {\bf 50}, 145 
(1994).
\bibitem{prlnoi} R. Roncaglia, L. Bonci, F. Izrailev, Bruce 
J. West, P. Grigolini, Phys. Rev. Lett. {\bf 73}, 802 (1994) 
\bibitem{frischat} E. Doron and S.D. Frischat, Phys. Rev. Lett {\bf 
75}, 3661 (1995); S.D. Frischat and E. Doron, Phys. Rev. E{\bf 57},
1421 (1998). 
\bibitem{lgw} M. Latka, P. Grigolini, B. J. West, Phys. Rev. E {\bf 
50}, 596 (1994); M. Latka, P. Grigolini, B. J. West, Phys. Rev. A 
{\bf 50}, 1071 (1994).
\bibitem{prenoi} L. Bonci, A. Farusi, P. Grigolini, R. Roncaglia, Phys. Rev. E
{\bf 58}, 5689 (1998).
\bibitem{hanggipreprint} S. Kohler, R. Utermann, P. Hanggi, T. 
Dittrich, Phys. Rev. E {\bf 58}, 7219 (1998).
\bibitem{fendrik}A.J. Fendrik and D.A. Wisniacki, Phys. Rev. E {\bf 
55}, 6507 (1997).
\bibitem{floq} Ya. B. Zel'dovich, Sov. Phys. JETP {\bf 24}, 1006 
(1967); V.I. Ritus, Sov. Phys. JETP {\bf 24}, 1041 (1967); J.H. 
Shirley, Phys. Rev. B {\bf 138}, 979 (1965).
\bibitem{r92} L.E. Reichl, {\it The transition to Chaos in 
Conservative Classical Systems: Quantum Manifestations} 
(Springer-Verlag, New York, 1992).
\bibitem{ll83} A.J. Lichtenberg and M.A. Lieberman, {\it Regular and 
Stochastic Motion} (Springer, New York, 1983). 
\bibitem{kam} A.N. Kolmogorov, Dok. Acad. Nauk SSSR {\bf 98}, 527 
(1954); V.I. Arnol'd, Usp. Mat. Nauk {\bf 18}, 13 (1963) 
[Russ. Math. Surv.{\bf 18}, 9 (1963)]; J. Moser, Nachr. Akad. Wiss.
G\"{o}ttingen {\bf 1962}, 1. 
\bibitem{idro}J. Zakrzewsky, D. Delande and A. Buchleitner, Phys. 
Rev. E {\bf 57}, 1458 (1997), K. Hornberger and A. Buchleitner, 
Europhys. Lett. {\bf 41}, 383 (1998).
\bibitem{ionization}G.N. Gibson, G. Dunne and K.J. Bergquist, Phys. Rev. Lett.
{\bf 81}, 2663 (1998).
\bibitem{split} M.D. Feit, J.A. Fleck, Jr. and A. Steiger, J. Comput. 
Phys. {\em 47}, 412 (1982).
\bibitem{kosloff}R. Kosloff and D. Kosloff, J. Comput. Phys. {\bf 
63}, 363 (1986).
\bibitem{husimi}W.H. Louisell, {\it Quantum Statistical Properties of 
Radiation} (Wiley, New York, 1990).  
\bibitem{semitunnel} H.J. Korsch, B. Mirbach and B. Schellhaass, J. Phys. 
A {\bf 30}, 1659 (1997).
\bibitem{breuer} H.P. Breuer and M. Holtaus, Ann. Phys. {\bf 211}, 249 (1991).
\bibitem{ebk} A. Einstein, Verh. Deutsch. Phys. Ges. Berlin {\bf 19},
82 (1917); L. Brillouin, J. Phys. Radium {\bf 7}, 353 (1926); 
J.B. Keller, Ann. Phys. (N.Y.) {\bf 4}, 180 (1958).
\bibitem{mp81} V. P. Maslov and M.V. Fedoriuk, {\it Semi-classical 
approximation in quantum mechanics}, D. Reidel Publishing Company, 
Dordrecht, Holland 1981.
\bibitem{a88} V.I. Arnold (Ed.), {\it Dynamic Systems III, 
Encyclopaedia of Mathematical Sciences}, Vol. 3, Springer-Verlag, 
Berlin, 1988. 
\bibitem{geiselband}R. Ketzmerick, K. Kruse, T. Geisel, Phys. Rev. 
Lett. {\bf 80}, 137 (1998).
\end{references}
\end{document}